\documentclass[onecolumn,showpacs,amsmath,amssymb,pra,letterpaper]{revtex4}

\usepackage{graphicx}

\begin{document}


\title{Exact distorted-wave approach to multiple-scattering theory
       for general potentials}

\author{D.~L.~Foulis}

\affiliation{European Synchrotron Radiation Facility, Bo\^{\i}te Postale 220,
             F-38043 Grenoble Cedex, France}

\date{\today}

\begin{abstract}
We present a new approach to real-space multiple-scattering theory for
molecules and clusters, based on the two-potential (distorted-wave)
Lippmann-Schwinger equation formalism. Our approach uses a recently
developed form [D. L. Foulis, Phys. Rev. A {\bf 70}, 022706 (2004)],
for the partial-wave expansions of the exact time-independent
single-particle Green function for a general potential, to solve exactly
the scattering problem for the distorting potential. The
multiple-scattering problem for the full multicenter molecular potential
is then developed along familiar lines, within a partition of space
consisting of non-overlapping atomic spheres, but relative to the
distorting potential. To achieve this some new general Green-function
re-expansion formulas are derived, as well as further developments of
our earlier partial-wave expansions. In passing, we make use of some
of our results to derive a complete solution to the
single-scattering problem for an arbitrary noncentral potential, and
investigate some of the algebraic properties of Gaunt coefficients
which arise frequently in our formulas. Based on the division of
the multicenter molecular potential into the non-singular
distorting potential and a remaining singular part
we develop explicitly the secular equations of our approach and
prove a result concerning the symmetry of the atomic matrices.
The new secular equations are similar in overall form to those of related
methods, but do not require any volume integrals. In our approach
the computational burden consists essentially of the solution
of the coupled radial Schr\"{o}dinger equations for each atomic center,
once in the actual atomic potential, and twice (giving the regular and
irregular solutions) in the distorting potential; followed by
integrals of combinations of these solutions over the atomic-sphere
surfaces. We comment on key aspects of the numerical implementation,
notably issues related to the choice of distorting potential.
We treat both continuum (scattering) states and bound states within
the same framework, and consider also the case of an outer sphere.
\end{abstract}



\maketitle

\newcommand{\vcr}{{\bf r}}
\newcommand{\vnr}{\widehat{\bf r}}
\newcommand{\vcs}{{\bf s}}
\newcommand{\vns}{\widehat{\bf s}}
\newcommand{\vct}{{\bf t}}
\newcommand{\vnt}{\widehat{\bf t}}
\newcommand{\vck}{{\bf k}}
\newcommand{\vnk}{\widehat{\bf k}}
\newcommand{\vcn}{{\bf n}}
\newcommand{\vnn}{\widehat{\bf n}}

\newcommand{\vcR}{{\bf R}}
\newcommand{\vnR}{\widehat{\bf R}}
\newcommand{\vcS}{{\bf S}}
\newcommand{\vnS}{\widehat{\bf S}}



\section{INTRODUCTION}

In previous work \cite*{Foulis04a} the author has derived partial-wave
expansions for the exact time-independent Green function of a general
noncentral one-electron potential in terms of matrix solutions of the
coupled radial Schr\"{o}dinger equations for that potential.
Although hopefully of interest in its own right to quantum-scattering
theorists, the motivation was, in fact, to provide a key result in
a new approach to multiple-scattering theory for
general potentials, which is intended to be less cumbersome,
more transparent, and more amenable to accurate
numerical implementation than alternative approaches.
In this article the full development of our new approach is presented.

We consider here multiple-scattering theory in the restricted sense of the
well known multiple-scattered-wave (MSW) theory of Johnson, Slater and
co-workers (see for example Ref. \cite{Johnso73a}) for bound states,
extended by others \cite{DilDeh74a} to continuum states and scattering,
as well as photoionization \cite{Na+al.80a}. It is expected, nevertheless,
that some of the results developed below will have wider application.
The MSW theory is itself an adaptation to finite clusters and molecules
of the famous KKR method from solid-state physics. Although we
concentrate here on methods for finite systems, where the practical
applications are varied and numerous, it is expected that the approach
of this present work may be adapted to the original case
of periodic systems.

As is well known, the main limitation for the original MSW and KKR methods
was the requirement of the so called ``muffin-tin'' (MT) approximation
for the potential. In this approximation space is partitioned into
disjoint atomic spheres and a remaining interstitial region.
The atomic potentials are taken to be spherically symmetric
and the interstitial potential constant, so that the motion of an 
electron in the potential may be described in terms of free propagation
between distinct atomic scattering events, for which the associated $T$
matrices are diagonal. There have been considerable efforts over many
years to extend these multiple-scattering methods to more general
potentials, which began to bear fruit towards the end of the 1980s,
and have led to some computational implementations.

One of the main reasons that so much work has been done in this area,
despite the fact that a variety of alternative, non-scattering methods
have become available (in certain fields of application),
is that few, if any, of these alternatives have all the
attractive features of the MSW/KKR approach. In particular, these are
real-space, non-basis-set methods which combine a
rapidly convergent partial-wave representation of multicenter
wavefunctions with an accurate treatment of atomic cores. Furthermore,
both bound and continuum states may be treated within the same framework,
making the methods applicable to a wide range of problems.
An attractive feature (of the original formalism at least), 
both from a conceptual and a practical point of view, is the
separation between geometry and scattering that is manifest in the
secular equations of the method.

The approach we shall present here has roots in the work of
Natoli, Benfatto, and Doniach~\cite{Na+al.86a} who developed a
particularly clear formulation of MSW theory for general potentials,
applied to molecules and clusters, drawing themselves on
the work of Beleznay and Lawrence~\cite{BelLaw68a}, and
Lloyd and Smith~\cite{LloSmi72a}. In their approach the atomic
$T$ matrices are generalized to describe scattering from a
non-spherical atomic potential, achieved by integrating the coupled
radial Schr\"{o}dinger equations in matrix form. The resulting
$T$ matrices then have non-zero off-diagonal elements.
The effect of a varying interstitial potential on the
electron scattering is taken into account via an extra term in
the MSW secular matrix, related to a $T$ matrix for the interstitial
potential.

This full-potential (FP) MSW method was implemented by the present
author and colleagues~\cite{Fo+al.90a} as the suite of computer codes
{\sc fpx}, and tested for analytically known bound-state and
continuum problems as well as realistic molecular
systems~\cite{Fo+al.90b} [in the context of {\it ab initio}
calculations in X-ray absorption fine-structure spectroscopy (XAFS)]. 
The codes have proved a useful pathfinder in exploring the benefits
of full-potential calculations and have produced results~\cite{FoPeSh95a}
which remain a benchmark~\cite{Rehr--02a} in the field.
They were designed to accommodate arbitrary
molecular geometries, although optimized for symmetric systems,
implementing the algebra of Ref.~\cite{Na+al.86a} completely,
except for the interstitial $T$ matrix which is calculated in the
Born approximation. They have also had a useful general role in
clarifying the relative importance of elements in the hierarchy of
approximations underlying XAFS calculations.

In practice, the {\sc fpx} codes have proved of limited applicability,
primarily because of the Born approximation for the interstitial $T$ matrix,
and the associated scaling behavior resulting from the volume integrals. 
A scheme to calculate the full interstitial $T$ matrix was developed by
W\"{a}stberg~\cite{Wastbe94a}, and implemented for bound states of
diatomic molecules. It is also a volume integral method relying on the
use of an accurate quadrature grid for the interstitial region.
Had development of the {\sc fpx} codes been continued his scheme would
undoubtedly have been incorporated.

Other schemes to extend MSW/KKR methods to non-MT potentials have been
developed, notably the ``cellular method'' in the context of band structure
calculations for periodic systems, proposed by Williams and
Morgan \cite{WilMor74a}, developed in a variety of forms and worked on
by many others. (To find a way into the literature of this approach
see Refs. \cite*{Foulis88a,Foulis04a,BuGoZh92a}
and further references therein.)

Despite a measure of progress, represented by the above approaches
(among others), it appears still to be the case that non-MT MSW/KKR
methods have been less widely taken up than might be expected.
It is true, in general, that all of these extensions are significantly
more complicated algebraically than the original MT-based methods,
and their numerical implementation correspondingly more challenging.
We shall not consider here in detail the reasons for the poor take-up
of these methods, although it seems likely to this author
that they are not unrelated to the latter observations.

It is perhaps not immediately obvious how to improve the situation.
The experience of the present author suggests that
the emphasis should be on reducing the computational burden
and, if not reducing algebraic complexity, at least increasing the
transparency of the theoretical approach. Also crucially important are
the details of the numerical implementation; although these will, of
course, be dictated to a great extent by the theory.
This present work is an attempt to develop a theory
with reduced computational requirements, which is also
well adapted to an accurate numerical implementation.

Our particular approach has grown from the simple notion, mentioned
in Ref. \cite{Foulis04a}, that a Green function better adapted 
than the free Green function, to the case of an electron propagating
in a varying potential, should play a central role in the theory.
This leads naturally to a consideration of the distorted-wave
Lippmann-Schwinger equation and how it may be adapted to the
multiple-scattering problem, and thus to the development of the
necessary mathematical machinery. As will be seen below the resulting
method retains the partition of space with distinct spherical atomic
volumes, but requires no volume integrals.
The computational burden is mainly the integration of coupled radial
Schr\"{o}dinger equations for each atomic center, and the evaluation of
surface integrals over the atomic spheres. 
It is hoped that the algebraic
development may be considered transparent and convincing, and
perhaps conceptually simpler than other approaches in this field.


\section{THEORY}

We consider a molecule or finite cluster of $N$ atoms with nuclear
positions given by the vectors ${\bf R}_i$ (for $i=1,\ldots,N$)
relative to some origin.
Associated with the molecule is a one-electron potential
$V(\vcr)$, arising perhaps in the context of Hartree-Fock or
density-functional calculations, of a multicenter character, i.e.,
with singularities at the nuclear sites, which tends to
zero (or some constant value) at infinity.
We are interested in solutions, both scattering (continuum) and
bound-state, of the Schr\"{o}dinger equation,
\begin{equation}
\left[ {\nabla}^2 + E - V(\vcr) \right] \psi(\vcr) = 0 \; ,
\label{eq:schreq}
\end{equation}
associated with the potential $V(\vcr)$.
(As in Ref.~\cite{Foulis04a} we use the Rydberg atomic units.)

As usual in multiple-scattering theory we shall partition the molecular
space as follows: each atomic center $i$ is enclosed in a spherical region
$\tau_i$ of radius $b_i$ centered on ${\bf R}_i$,
\begin{equation}
\tau_i \equiv \left\{ \vcr \in {\mathbb R}^3 :
\left| \vcr - \vcR_i \right| < b_i \right\} \, ,
\end{equation}
taken to be mutually disjoint (non-overlapping), so that
$b_i + b_j \leq \left| \vcR_i - \vcR_j \right|$ for all $i$ and $j$
not equal. The surface of atomic sphere $i$ will be denoted by
$\partial \tau_i$, i.e.,
\begin{equation}
\partial \tau_i \equiv \left\{ \vcr \in {\mathbb R}^3 :
\left| \vcr - \vcR_i \right| = b_i \right\} \, .
\end{equation}
We shall consider later a modified partition which includes also
an outer sphere $\tau_0$ of radius $b_0$ centered on some position
$\vcR_0$, which contains all the atomic spheres, i.e.,
\begin{equation}
\tau_0 \equiv \left\{ \vcr \in {\mathbb R}^3 :
\left| \vcr - \vcR_0 \right| \leq b_0 \right\} \, ,
\end{equation}
such that $\tau_i \subset \tau_0$ for all $i$. In this case the
region outside the outer sphere, denoted by $\overline{\tau_0}$
(the complement of $\tau_0$), will be treated in effect as an atomic center.


\subsection{The distorted-wave Lippmann-Schwinger equation}

In the usual multiple-scattering methods derivation of
the central equations frequently begins
with the Lippmann-Schwinger equation for scattering
states $\psi(\vcr)$ in a multicenter potential $V(\vcr)$,
\begin{equation}
\psi(\vcr) = \phi_0(\vcr) + \int G_0^+(\vcr,\vcs) V(\vcs)
\psi(\vcs) \; d^3\!s \, ,
\label{eq:lippschw}
\end{equation}
where $\phi_0(\vcr)$ is a plane wave
and $G_0^+(\vcr,\vcs)$ is the time-independent free Green function
with outgoing-wave boundary conditions.
(An integral in which the region is not specified should be taken over the
whole range of the variable of integration; e.g., in the foregoing equation
this is therefore over all space.)

The key to the approach we present here is the splitting of our
multicenter molecular potential into two parts: a finite non-singular
potential $V_I(\vcr)$ which coincides with $V(\vcr)$ in the interstitial
region; and a remaining part $V_A(\vcr)$
which is zero in the interstitial region,
but contains atomic potential singularities. Thus,
\begin{equation}
V(\vcr) = V_I(\vcr) + V_A(\vcr) \, ,
\label{eq:potsplit}
\end{equation}
with
\begin{equation}
V_I(\vcr) \equiv \left\{
   \begin{array}{lll}
      V_{IA}(\vcr)   &  &  \mbox{if $\vcr \in \bigcup_{i=1}^N \tau_i$} \\
      V(\vcr)   &  &  \mbox{otherwise}
   \end{array}
\right. \; .
\end{equation}
It is not necessary for our immediate purpose to define $V_{IA}(\vcr)$ other
than that it lead to a non-singular, well behaved, and finite $V_I(\vcr)$.
One possible choice is to set it to be identically zero. Alternatively,
one might choose a form which joins continuously or smoothly to
$V(\vcr)$ at the atomic-sphere surfaces. Clearly the actual choice will
be significant when one considers the numerical implementation of the
methods we present here and is
a point which we shall consider in more detail later on.

Having made this split of the potential our approach to the
multiple-scattering problem will be via ``distorted-wave''
theory~\cite{Newton66a}, otherwise known
as the case of additive potentials, or the two-potential
formula~\cite{RodTha67a,Taylor72a},
which we shall use here without approximation. In this theory,
instead of taking the zeroth-order hamiltonian
to be the kinetic energy operator,
\begin{equation}
H_0 \equiv - \nabla^2 \, ,
\end{equation}
and the scattering potential to be $V(\vcr)$, leading to the usual
Lippmann-Schwinger equation~(\ref{eq:lippschw}), we take it to be
\begin{equation}
H_I \equiv - \nabla^2 + V_I \, ;
\end{equation}
with the full hamiltonian being
\begin{equation}
H \equiv - \nabla^2 + V \, .
\end{equation}
We may then obtain the desired outgoing-wave scattering solutions
$\psi^+$ of the
full hamiltonian by solving the Lippmann-Schwinger equation
\begin{equation}
\psi^+(\vcr) = \chi^+(\vcr) + \int G_I^+(\vcr,\vcs) V_A(\vcs)
\psi^+(\vcs) \; d^3\!s \, ,
\label{eq:lippschwdw}
\end{equation}
where $G_I^+$ is the Green function relative to the potential $V_I$,
with outgoing-wave boundary conditions, and the ``distorted wave''
$\chi^+(\vcr)$ satisfies the Lippmann-Schwinger
equation for the ``distorting'' potential $V_I(\vcr)$, i.e.,
\begin{equation}
\chi^+(\vcr) = \phi_0(\vcr) + \int G_0^+(\vcr,\vcs) V_I(\vcs)
\chi^+(\vcs) \; d^3\!s \, .
\label{eq:lippschwdwfn}
\end{equation}
As is well known (see Ref.~\cite{RodTha67a} page 132)
this may be replaced by the explicit form
\begin{equation}
\chi^+(\vcr) = \phi_0(\vcr) + \int G_I^+(\vcr,\vcs) V_I(\vcs)
\phi_0(\vcs) \; d^3\!s \, ,
\label{eq:lippschwdwfnex}
\end{equation}
from which, if one knows the explicit form of $G_I^+$, one may evaluate
$\chi^+$ directly. However, our previous work \cite{Foulis04a} gives us
precisely this, as we shall see in detail below.


\subsection{Partial-wave Green-function expansions}

In Ref.~\cite{Foulis04a} the author has derived an expression for
the partial-wave expansions of the exact time-independent Green functions,
at positive and negative energies, for an arbitrary single-particle
potential $V(\vcr)$ (within certain limitations).
Since the results of this section will be completely general
we shall, until further notice, take $V(\vcr)$ not to be
our multicenter potential, introduced earlier. In fact, the results
of Ref.~\cite{Foulis04a}, and some immediate consequences which we
develop below, will be applied to the more well behaved
part $V_I(\vcr)$ of the multicenter potential, providing us with
explicit forms for $G_I^+(\vcr,\vcs)$, therefore allowing us
to attack the multicenter problem via the integral equations
(\ref{eq:lippschwdw}) and (\ref{eq:lippschwdwfnex}).

The Green functions that we require are solutions of the equation
\begin{equation}
\left[ {\nabla_{\vcr}}^2 + E - V(\vcr) \right] G(\vcr,\vcs) =
\delta^3(\vcr - \vcs) \; ,
\label{eq:gdef}
\end{equation}
together with the appropriate boundary conditions.
The potential $V({\vcr})$ is given as a well behaved
convergent expansion in spherical harmonics
about some point which we take as the origin of coordinates, so that
\begin{equation}
V(\vcr) \equiv \sum_L v_L(r) Y_L(\vnr) \; ,
\label{eq:vshx}
\end{equation}
where the compound index $L$ represents, as usual, the pair $(l,m)$ and
the summation over $L$ is given by
\begin{equation}
\sum_L \equiv \sum_{l=0}^{\infty} \sum_{m=-l}^{l} \; .
\end{equation}
Again to simplify the algebra we shall make use of real spherical harmonics
(where the index $m$ is negative for sine and positive or zero for cosine
spherical harmonics). As is well known they may be obtained from complex
spherical harmonics by a unitary transformation which preserves
the orthonormality property and the addition theorem.

The Green functions may be expressed in terms of solutions
of the coupled radial Schr\"{o}dinger equations.
These may derived from Eq.~(\ref{eq:schreq}) by considering the
solution of Schr\"{o}dinger's equation as a spherical-harmonic expansion,
\begin{equation}
\psi(\vcr) \equiv \sum_L \phi_L(r) Y_L(\vnr) \; ,
\end{equation}
with which one quickly finds that
\begin{equation}
\left[ \frac{1}{r^2} \left( \frac{\partial}{\partial r} \left( r^2
\frac{\partial}{\partial r} \right) \right)
- \frac{l(l+1)}{r^2} + E
\right] \phi_{L}(r)
 - \sum_{L'} w_{L L'}(r) \phi_{L'}(r) = 0 \; ;
\label{eq:coupradschreq}
\end{equation}
where we have used spherical-harmonic orthonormality and
defined the potential matrix elements
\begin{equation}
w_{L L'}(r) \equiv \sum_{L''} I(L,L',L'') v_{L''}(r)
\label{eq:potmx}
\end{equation}
(with the Gaunt coefficients for real spherical harmonics given by
\begin{equation}
I(L,L',L'') \equiv \int Y_L(\vnt) Y_{L'}(\vnt) Y_{L''}(\vnt) \;
d\Omega_{\vct}
\label{eq:gauntdef}
\end{equation}
which are manifestly symmetric under any permutation of the indices
$L$, $L'$, and $L''$).
It is
obvious that the quantities $w_{L L'}$ are real, for a real potential, and
symmetric in the indices.

The general form of the Green function may be now written
as follows \cite*{Foulis04a}:
\begin{eqnarray}
G(\vcr,\vcs) & = & \sum_{L L_1 L_2 L'} \left[
p_{L L_1}(r) \left\{ M^{-1} \right\}_{L_2 L_1} q_{L' L_2}(s) \theta(s-r)
\right. \nonumber \\
 & & \left.
+ q_{L L_1}(r) \left\{ M^{-1} \right\}_{L_1 L_2} p_{L' L_2}(s) \theta(r-s)
\right] Y(\vnr)_L Y(\vns)_{L'} \; ,
\label{eq:gfinalsbs}
\end{eqnarray}
where the $p_{L L'}(r)$ constitute a set of real linearly independent
vector solutions to the radial
Schr\"{o}dinger equations (\ref{eq:coupradschreq}), indexed by $L'$,
and regular at $r = 0$;
the $q_{L L'}(r)$ constitute similarly a set of vector solutions,
regular at infinity, whose boundary conditions at this limit determine
which Green function we have; and $M_{L L'}$
is a constant matrix derived from
the Wronskian of these two sets of functions, as we describe below.
Since the potential will, in general, have non-zero higher multipole
components, the potential matrix $w_{L L'}$
will have non-zero off-diagonal
elements. Consequently the solution matrices will not be diagonal.

It is interesting to note that the problem of an explicit form for the
partial-wave expansions of the Green function of a noncentral potential
appears to have been rarely treated. One example we have
found is Ref.~\cite{BuGoZh92a}, where the authors consider the problem
in their Appendix, using the phase-functional formalism with
particular boundary conditions. They develop an expression without an
explicit matrix $M_{L L'}$, which seems not readily adapted to the
further development we shall present later.

It is useful before proceeding further to introduce some notation. In
particular we represent our infinite matrices and vectors by sans-serif
letters. Therefore we may write equations without explicit reference to
indices and summations. Thus, in an obvious way, our set of radial matrix
solutions may be represented by ${\sf p}(r)$ and ${\sf q}(r)$, and the
constant matrix by ${\sf M}$.

Our general result may be now written more clearly as
\begin{equation}
G(\vcr,\vcs) = {\sf Y}(\vnr)^T \left[
{\sf p}(r) \left( {\sf M}^{-1} \right)^{T} {\sf q}(s)^T \theta(s-r)
+ {\sf q}(r) {\sf M}^{-1} {\sf p}(s)^T \theta(r-s)
\right] {\sf Y}(\vns) \; ,
\label{eq:gfinal}
\end{equation}
where ${\sf Y}(\vnr)$ is a column vector with elements $Y_L(\vnr)$,
the superscript $T$ denotes the matrix transpose, and, as was shown in
Ref.~\cite{Foulis04a} using a well known argument, we have
\begin{equation}
{\sf p}'(r)^T {\sf q}(r)
- {\sf p}(r)^T {\sf q}'(r)  = - \frac{1}{r^2} {\sf M} \; .
\label{eq:pqwronski}
\end{equation}
In fact the argument is valid for any pair of matrix solutions, leading to
a different constant matrix in each case. When the two matrix solutions
are the same, further argument shows, at least for ${\sf p}$ and ${\sf q}$
individually, that the constant matrix is zero, so that
\begin{equation}
{\sf p}(r)^T {\sf p}'(r) = {\sf p}'(r)^T {\sf p}(r) \; ,
\label{eq:ppsym}
\end{equation}
and
\begin{equation}
{\sf q}(r)^T {\sf q}'(r) = {\sf q}'(r)^T {\sf q}(r) \; ,
\label{eq:qqsym}
\end{equation}
provided that, in the latter case, ${\sf q}$ tends to a diagonal form
in the asymptotic limit, which will usually be the case if
the potential tends to zero quickly enough asymptotically.
These two equations would have been automatically true in the
case of diagonal matrices when the potential is spherically symmetric.

In the course of the derivation of our main result other relations
between these matrices arise. In particular, we have
\begin{equation}
{\sf p}(r) \left( {\sf M}^{-1} \right)^{T} {\sf q}(r)^T =
{\sf q}(r) {\sf M}^{-1} {\sf p}(r)^T \; ,
\label{eq:gpartcont}
\end{equation}
and
\begin{equation}
{\sf q}'(r) {\sf M}^{-1} {\sf p}(r)^T -
{\sf p}'(r) \left( {\sf M}^{-1} \right)^{T} {\sf q}(r)^T =
\frac{1}{r^2} \,  {\sf I} \; ,
\label{eq:gpartinhom}
\end{equation}
both of which will be useful later on.

The choice of particular Green function depends on the boundary
conditions, which are imposed on the asymptotic behavior
of ${\sf q}$. For positive energies, $E>0$ (with $k>0$ such that
$E=k^2$), and Green functions suitable for scattering problems, we
have two choices for ${\sf q}$. Firstly, the choice of outgoing-wave
boundary conditions and the retarded Green function is obtained with
${\sf q}^{(+)}$, defined to be the solutions of the coupled radial
Schr\"{o}dinger equations (\ref{eq:coupradschreq}) which approach
asymptotically, as $r \rightarrow \infty$, 
the corresponding free-electron solutions
\begin{equation}
{\sf q}^{(+)}_f(r) \equiv -i {\sf h}^+(kr) \, ,
\label{eq:qplusfree}
\end{equation}
where the diagonal spherical Hankel-function matrix has elements
given by
\begin{equation}
h^+_{L L'}(x) \equiv h^+_l(x) \delta_{L L'} \, .
\end{equation}
We define here similarly spherical Bessel- and Neumann-function
matrices ${\sf j}$ and ${\sf n}$, respectively, by
\begin{equation}
j_{L L'}(x) \equiv j_l(x) \delta_{L L'}
\end{equation}
and
\begin{equation}
n_{L L'}(x) \equiv n_l(x) \delta_{L L'} \, ;
\end{equation}
so that
\begin{equation}
{\sf h}^+(x) = {\sf j}(x) + i {\sf n}(x) \, .
\end{equation}
It is useful to define the diagonal unitary matrices
$\xi$ and $\eta$ by
\begin{eqnarray}
[\xi]_{L L'} \equiv i^l \delta_{L L'} & \mbox{and} &
[\eta]_{L L'} \equiv (-1)^l \delta_{L L'} \, ,
\label{eq:xietadefs}
\end{eqnarray}
from which we may see that
\begin{eqnarray}
\xi^2 = \eta & \mbox{and} & \xi^4 = \eta^2 = {\sf I} \, .
\label{eq:xietaprops}
\end{eqnarray}
With this notation we may write the famous plane-wave expansion as
\begin{equation}
e^{i \vck \cdot \vcr} = 4 \pi \; {\sf Y}(\vnk)^T \; {\sf j}(kr)
\; \xi \; {\sf Y}(\vnr) \, .
\label{eq:planewavexp}
\end{equation}

The choice of incoming-wave
boundary conditions and the advanced Green function is obtained with
${\sf q}^{(-)}$, defined to be the complex conjugate of ${\sf q}^{(+)}$;
\begin{equation}
{\sf q}^{(-)}(r) = {\sf q}^{(+)}(r)^* \, .
\end{equation}
The constant matrices ${\sf M}$ from Eq.~(\ref{eq:pqwronski}),
for the retarded and advanced Green functions, are denoted
${\sf M}_+$ and ${\sf M}_-$ respectively. It is easily seen from
Eq.~(\ref{eq:pqwronski}), since ${\sf p}$ is real, that
\begin{equation}
{\sf M}_+ = {{\sf M}_-}^* \; .
\end{equation}
By the argument which led to Eq.~(\ref{eq:pqwronski}) we have a similar
equation for the Wronskian between ${\sf q}^{(+)}$
and ${\sf q}^{(-)}$, however,
the constant matrix may be evaluated by considering the limit as
$r \rightarrow \infty$, and making use of the well known relation
for spherical Bessel and Neumann functions \cite{AbrSte65a},
\begin{equation}
j'_l(x) n_l(x) - j_l(x) n'_l(x) = - \frac{1}{x^2} \; .
\label{eq:bessneuwr}
\end{equation}
Thus we find that
\begin{equation}
{{\sf q}^{(+)}}'(r)^T {\sf q}^{(-)}(r)
- {\sf q}^{(+)}(r)^T {{\sf q}^{(-)}}'(r)  = \frac{2 i}{k r^2} {\sf I} \; .
\label{eq:qpqmwronski}
\end{equation}

When the potential vanishes we have, of course, the free-electron case
and can take ${\sf p}$ and ${\sf q}$ to be, respectively,
\begin{equation}
{\sf p}_f(r) \equiv {\sf j}(kr)
\end{equation}
and ${\sf q}^{(+)}_f(r)$ from Eq.~(\ref{eq:qplusfree}), leading to
constant matrix
\begin{equation}
{\sf M}_{+ f} \equiv (1/k) {\sf I} \, ,
\end{equation}
and free retarded Green function
\begin{equation}
G^+_0(\vcr,\vcs) = -i k \sum_L  \left[ j_l(k r) h_l^+(k s)
\theta(s-r) + h_l^+(k r) j_l(k s) \theta(r-s) \right]
Y_L(\vnr) Y_L(\vns) \; ,
\end{equation}
with its famous closed-form representation
\begin{equation}
G^+_0(\vcr,\vcs) = - \frac{e^{i k  \left| \vcr - \vcs \right|}}
{4 \pi \left| \vcr - \vcs \right|} \; .
\end{equation}

For negative energies, $E<0$ (with $\kappa>0$ such that
$E=\kappa^2$), we have only one acceptable set of functions
${\sf q}_f$ regular at infinity, which tends asymptotically
to a diagonal exponentially decaying form.
In general, the singularity of the
Green function at a bound-state energy comes about through the
singularity of the constant matrix ${\sf M}_f$ at that energy.

In the free-electron case
we can take ${\sf p}$ and ${\sf q}$ to be, respectively,
\begin{equation}
{\sf p}_f(r) \equiv {\sf i}(\kappa r)
\label{eq:pnegfree}
\end{equation}
and
\begin{equation}
{\sf q}_f(r) \equiv {\sf k}^+(\kappa r),
\label{eq:qnegfree}
\end{equation}
where, as above, we have defined matrices for the usual modified
spherical Bessel and Hankel functions as
\begin{equation}
i_{L L'}(x) \equiv i_l(x) \delta_{L L'}
\end{equation}
and
\begin{equation}
{k^+}_{L L'}(x) \equiv {k^+}_l(x) \delta_{L L'} \, .
\end{equation}
These definitions lead to constant matrix
\begin{equation}
{\sf M}_{f} \equiv - (1/\kappa) {\sf \eta} \, ,
\end{equation}
and Green function
\begin{equation}
G_0(\vcr,\vcs) = - \kappa \sum_L (-1)^l \left[ i_l(\kappa r) k_l^+(\kappa s)
\theta(s-r) + k_l^+(\kappa r) i_l(\kappa s) \theta(r-s) \right]
Y_L(\vnr) Y_L(\vns) \; ,
\end{equation}
with closed-form representation
\begin{equation}
G_0(\vcr,\vcs) = - \frac{e^{- \kappa  \left| \vcr - \vcs \right|}}
{4 \pi \left| \vcr - \vcs \right|} \; .
\end{equation}

To finish this section we note that the standard
symmetries of the general Green
function (see for example \cite{Prugov71a} and \cite{Schiff68a}),
\begin{equation}
G^+(\vcr,\vcs)  = G^-(\vcs,\vcr)^*
\end{equation}
(stemming from the reality of $V(\vcr)$), and,
\begin{equation}
G^+(\vcr,\vcs)  = G^+(\vcs,\vcr) \; ,
\label{eq:gvarsym}
\end{equation}
may be be verified from the main formula~(\ref{eq:gfinal}).


\subsubsection{Further developments}

In the foregoing section we have reproduced the main results from
Ref.~\cite{Foulis04a} together with some useful incidental formulas,
and set up some notation which will be used in the remainder of this work.
In this section we develop some immediate consequences of these results.
Returning to the positive-energy scattering situation, we derive first an
important relation between the radial Schr\"{o}dinger equation solutions
${\sf p}$, ${\sf q}^{(+)}$, and ${\sf q}^{(-)}$.

Since ${\sf q}^{(+)}$ and ${\sf q}^{(-)}$ are the two linearly independent
radial-solution matrices regular at infinity, then any other solution must
be expressible as a linear combination of them. This applies therefore to
${\sf p}$. (We note that all three sets of solutions are well defined at
all radii, determined by inward or outward integration of the coupled radial
equations from suitable boundary conditions; except for ${\sf q}^{(+)}$
and ${\sf q}^{(-)}$ which may be singular at $r=0$.) Thus, given any
constant vector ${\sf X}$ (with elements $X_L$), there must exist constant
vectors ${\sf Z}^+$ and ${\sf Z}^-$ such that, for general $r$,
\begin{equation}
{\sf p}(r){\sf X} = {\sf q}^{(+)}(r){\sf Z}^+ 
+ {\sf q}^{(-)}(r){\sf Z}^- \; .
\label{eq:pqpqm}
\end{equation}
Evidently we may differentiate both sides with respect to $r$ to get a
similar relation between their derivatives. Thus we may form the matrix
Wronskians of both sides with
${\sf q}^{(+)}$, ${\sf q}^{(-)}$, and
${\sf p}$, successively, and make use of Eqs.
(\ref{eq:pqwronski}), (\ref{eq:ppsym}), (\ref{eq:qqsym}),
and (\ref{eq:qpqmwronski}),
to obtain, respectively,
(with some minor rearrangement) three new relations:
\begin{equation}
{\sf Z}^- = - \left( \frac{i k}{2} \right) {{\sf M}_+}^T {\sf X} \; ,
\nonumber
\end{equation}
\begin{equation}
{\sf Z}^+ =   \left( \frac{i k}{2} \right) {{\sf M}_-}^T {\sf X} \; ,
\nonumber
\end{equation}
and
\begin{equation}
{\sf M}_+ {\sf Z}^+ + {\sf M}_- {\sf Z}^- = 0 \; .
\nonumber
\end{equation}
(Note that ${\sf Z}^+$ and ${\sf Z}^-$ are complex conjugate only if
${\sf X}$ is real.) We may use the first two of these relations to
eliminate ${\sf Z}^+$ and ${\sf Z}^-$ from our original equation
(\ref{eq:pqpqm}), and also from the third above, to obtain, noting
that ${\sf X}$ is arbitrary,
\begin{equation}
{\sf p}(r) = \frac{i k}{2} \left[ {\sf q}^{(+)}(r){{\sf M}_-}^T
- {\sf q}^{(-)}(r){{\sf M}_+}^T \right] \; ,
\label{eq:pqpmmqmmp}
\end{equation}
and
\begin{equation}
{\sf M}_+ {{\sf M}_-}^T = {\sf M}_- {{\sf M}_+}^T \; .
\label{eq:mmmptsym}
\end{equation}
From the latter equation we see that the combination
${\sf M}_- {{\sf M}_+}^T$ is symmetric. Since ${\sf M}_-$ and
${{\sf M}_+}$ are complex conjugate, then ${\sf M}_- {{\sf M}_+}^T$ is
hermitian. Therefore it is also real.

We note, in the positive-energy case, that the
relations (\ref{eq:gpartcont}) and (\ref{eq:gpartinhom}),
between combinations of the radial functions and their derivatives,
are valid separately for ${\sf q}^{(+)}$ and ${\sf q}^{(-)}$.
It will be useful for later reference to prove a further, similar
relation. So, given ${\sf p}$ and ${\sf q}$,
let us define ${\sf y}$ and ${\sf z}$ by
\begin{eqnarray}
{\sf y}(r) \equiv r {\sf p}(r) & \mbox{and} & {\sf z}(r) \equiv r
{\sf q}(r) \; .
\label{eq:yzdefs}
\end{eqnarray}
Define also ${\sf u}(r)$, as in Ref.~\cite{Foulis04a}, by
\begin{equation}
u_{L L'}(r) \equiv w_{L L'}(r) + \frac{l(l+1)}{r^2} \delta_{L L'}
- E \delta_{L L'} \; ,
\end{equation}
noting definition (\ref{eq:potmx}) and the radial Schr\"{o}dinger equation
(\ref{eq:coupradschreq}), from which it may be easily seen that
\begin{eqnarray}
{\sf y}''(r) = {\sf u}(r) {\sf y}(r) & \mbox{and} &
{\sf z}''(r) = {\sf u}(r) {\sf z}(r) \; .
\label{eq:yzschreq}
\end{eqnarray}
With these definitions it is not difficult to show, using Eqs.
(\ref{eq:gpartcont}) and (\ref{eq:gpartinhom}), that
\begin{equation}
{\sf z}'(r) {\sf M}^{-1} {\sf y}(r)^T -
{\sf y}'(r) \left( {\sf M}^{-1} \right)^{T} {\sf z}(r)^T =
{\sf I} \; .
\nonumber
\end{equation}
Taking the derivative of this with respect to $r$, eliminating second
derivatives with Eq.~(\ref{eq:yzschreq}), and using
Eq.~(\ref{eq:gpartcont}) to make cancellations, leads to
\begin{equation}
{\sf z}'(r) {\sf M}^{-1} {\sf y}'(r)^T =
{\sf y}'(r) \left( {\sf M}^{-1} \right)^{T} {\sf z}'(r)^T \; .
\nonumber
\end{equation}
If we now substitute for ${\sf y}$ and ${\sf z}$ from
Eq.~(\ref{eq:yzdefs}), and make some cancellations with
Eqs.~(\ref{eq:gpartcont}) and (\ref{eq:gpartinhom}), then we obtain
our result:
\begin{equation}
{\sf q}'(r) {\sf M}^{-1} {\sf p}'(r)^T =
{\sf p}'(r) \left( {\sf M}^{-1} \right)^{T} {\sf q}'(r)^T \; .
\label{eq:gpartd1d1}
\end{equation}

With our most important result (\ref{eq:pqpmmqmmp}) from earlier in this
section it is possible to eliminate ${\sf p}$ from the set of relations
(\ref{eq:gpartcont}), (\ref{eq:gpartinhom}), and (\ref{eq:gpartd1d1}), to
obtain a similar set involving only ${\sf q}^{(+)}$ and ${\sf q}^{(-)}$.
To carry this out we need first to note that Eq.~(\ref{eq:mmmptsym})
may be rearranged as
\begin{equation}
{{\sf M}_-}^T \left( {\sf M}_+^{-1} \right)^{T} 
= {\sf M}_+^{-1} {\sf M}_- \; ,
\label{eq:mmmpsym}
\end{equation}
showing incidentally that the combination
${{\sf M}_-}^T \left( {\sf M}_+^{-1} \right)^{T}$, which will prove
important in the next section, is a symmetric matrix.
One may proceed straightforwardly now to obtain, respectively,
\begin{equation}
{\sf q}^{(+)}(r) {\sf q}^{(-)}(r)^T =
{\sf q}^{(-)}(r) {\sf q}^{(+)}(r)^T \; ,
\end{equation}
\begin{equation}
{{\sf q}^{(+)}}'(r) {\sf q}^{(-)}(r)^T
- {{\sf q}^{(-)}}'(r) {{\sf q}^{(+)}}(r)^T  = 
- \frac{2 i}{k r^2} {\sf I} \; ,
\end{equation}
which may be compared with the Wronskian (\ref{eq:qpqmwronski}), and
\begin{equation}
{{\sf q}^{(+)}}'(r) {{\sf q}^{(-)}}'(r)^T =
{{\sf q}^{(-)}}'(r) {{\sf q}^{(+)}}'(r)^T \; .
\end{equation}


\subsection{Single scattering and the distorted wave}

Our approach to the multiple-scattering problem is via the distorted-wave
Lippmann-Schwinger equation (\ref{eq:lippschwdw}). This requires, of course,
that we have an explicit expression for the distorted wave $\chi^+$;
which means, in effect, a complete solution to the single-scattering problem
for the distorting potential $V_I$. The fact that we have a formula
for the Green function $G_I^+$, from our earlier results, enables precisely
that. Since the results of this section can be applied to any general,
noncentral potential (within the limitations discussed in
Ref.~\cite{Foulis04a}), we drop the subscript $I$, and continue to use $V$
to represent an arbitrary potential.

To find $\chi^+$ then, we use the explicit version
(\ref{eq:lippschwdwfnex}) of the Lippmann-Schwinger equation
\begin{equation}
\chi^+(\vcr) = \phi_0(\vcr) + \int G^+(\vcr,\vcs) V(\vcs)
\phi_0(\vcs) \; d^3\!s \, ,
\end{equation}
with
\begin{equation}
\phi_0(\vcr) \equiv e^{i \vck \cdot \vcr} \; .
\end{equation}
Now because of the inhomogeneous equation (\ref{eq:gdef})
satisfied by $G^+$, and the symmetry (\ref{eq:gvarsym}),
we can replace the product $G^+ V$ in the above equation
with $\left( {\nabla_{\vcs}}^2 + E \right) G^+(\vcr,\vcs)
- \delta^3(\vcr - \vcs)$. Canceling the inhomogeneous part with the
delta-function term we therefore obtain
\begin{equation}
\chi^+(\vcr) = \int \phi_0(\vcs) \left( {\nabla_{\vcs}}^2 + E \right)
G^+(\vcr,\vcs) \; d^3\!s \; .
\label{eq:chipls}
\end{equation}
Since $\phi_0$ satisfies the free-particle Schr\"{o}dinger equation,
\begin{equation}
\left( {\nabla}^2 + E \right) \phi_0(\vcr) = 0 \; ,
\end{equation}
we can add an extra term to the right-hand side (rhs) to get
\begin{equation}
\chi^+(\vcr) = \int \left[ \phi_0(\vcs) \left( {\nabla_{\vcs}}^2 + E \right)
G^+(\vcr,\vcs) - G^+(\vcr,\vcs) \left( {\nabla_{\vcs}}^2 + E \right)
\phi_0(\vcs) \right] \; d^3\!s \; .
\label{eq:chiplssym}
\end{equation}

To transform the volume integral into a surface integral, we now wish
to use Green's theorem in the form (see Ref.~\cite{Na+al.86a})
\begin{equation}
\int_{T} \left[ P \left( {\nabla}^2 + E \right) Q
- Q \left( {\nabla}^2 + E \right) P \right] \; d\tau
= \int_{\partial T} \left( P {\nabla} Q - Q {\nabla} P \right)
\cdot \vnn \; d\sigma \; ,
\label{eq:greenthm}
\end{equation}
where $\partial T$ is the surface enclosing the volume $T$, and $\vnn$
is the outward-pointing normal to the surface. It should be noted here
that, since $\phi_0$ is not a proper normalizable vector
in the Hilbert space of physical states, the volume integral in
Eq.~(\ref{eq:chiplssym}) is not zero. When transformed by Green's
theorem to a surface integral, the ``surface at infinity'' provides
a non-zero contribution which is the quantity we seek.

To find this quantity we first consider the volume integral of
Eq.~(\ref{eq:chiplssym}) as being over a spherical region with
large radius $R$, and then allow $R$ to go to infinity.
Using Green's theorem then leads to
\begin{equation}
\chi^+(\vcr) = \lim_{R \rightarrow \infty} \int \left.
\left\{ \phi_0(\vcs) \frac{\partial}{\partial s} \left[
G^+(\vcr,\vcs) \right] - G^+(\vcr,\vcs)
\frac{\partial \phi_0(\vcs)}{\partial s} \right\} \right|_{s=R}
R^2 \; d\Omega_{\vcs} \; ,
\label{eq:chipgt}
\end{equation}
where we have made use of the fact that
$\vnn \cdot \nabla_{\vcs} = {\partial} / {\partial s}$. Now if $R$
is larger than $r$, then we can take the Green function to consist
of only the first term of Eq.~(\ref{eq:gfinal}),
with the choice of outgoing waves ${\sf q}^{(+)}$.
The plane wave $\phi_0$ may also be expressed as the expansion
(\ref{eq:planewavexp}). Substituting both expressions into the
above equation for $\chi^+$ leads, with some rearrangement, to
\begin{equation}
\chi^+(\vcr) =
4 \pi \; {\sf Y}(\vnr)^T {\sf p}(r) \left( {\sf M}_+^{-1} \right)^{T}
\left\{ \lim_{R \rightarrow \infty} R^2 \left[
{{\sf q}^{(+)}}'(R)^T {\sf j}(k R)
- {{\sf q}^{(+)}}(R)^T k {\sf j}'(k R)
\right] \right\} \xi \; {\sf Y}(\vnk) \, ,
\label{eq:chipprelim1}
\end{equation}
where we have used the orthonormality of spherical harmonics,
\begin{equation}
\int {\sf Y}(\vns) {\sf Y}(\vns)^T \; d\Omega_{\vcs} = {\sf I} \; ,
\end{equation}
to remove the angular integral. To evaluate the limit of the Wronskian
expression we make use of the asymptotic form of ${\sf q}^{(+)}$
and invoke the formula (\ref{eq:bessneuwr}). The result for the expression
inside the braces is easily found to be $(1/k) {\sf I}$.
Our final result is then
\begin{equation}
\chi^+(\vcr) = ( 4 \pi / k ) \;
{\sf Y}(\vnr)^T {\sf p}(r)
\left( {\sf M}_+^{-1} \right)^{T}
\xi \; {\sf Y}(\vnk) \; ,
\label{eq:chip}
\end{equation}
the general form of the scattering state arising from a plane wave in a
noncentral potential.


\subsubsection{Single scattering: complete solution}

Our equation (\ref{eq:chip}) contains all the information required for
the solution of the single-scattering problem for a general potential,
but is not usable for this purpose in its present form since it contains
the set of radial solutions ${\sf p}$ regular at $r=0$. From the
scattering point of view we are more interested in the asymptotic region.
With this in mind we remove ${\sf p}$ by using the formula
(\ref{eq:pqpmmqmmp}). This gives us straightforwardly
\begin{equation}
\chi^+(\vcr) =
2 \pi i \; {\sf Y}(\vnr)^T
\left[ {\sf q}^{(+)}(r) {\sf \hat{A}} - {\sf q}^{(-)}(r) \right]
\xi \; {\sf Y}(\vnk) \; ,
\end{equation}
where we define
\begin{equation}
{\sf \hat{A}} \equiv {{\sf M}_-}^T \left( {\sf M}_+^{-1} \right)^{T} \; ,
\label{eq:defofahat}
\end{equation}
the symmetric matrix that we encountered earlier. We note therefore that
\begin{equation}
{\sf \hat{A}}^{\dagger} = {\sf \hat{A}}^* =
{{\sf M}_+}^T \left( {\sf M}_-^{-1} \right)^{T} = {\sf \hat{A}}^{-1} \; ;
\end{equation}
i.e., ${\sf \hat{A}}$ is unitary.

Since the asymptotic behavior of $\chi^+$ is more interesting for the
single-scattering problem, we consider the form of $\chi^+$
at large distances where, by construction, ${\sf q}^{(+)}$ and
${\sf q}^{(-)}$ approach their free-particle counterparts.
Thus, at large $r$,
\begin{equation}
\chi^+(\vcr) \sim
2 \pi i \; {\sf Y}(\vnr)^T
\left[ {\sf q}_f^{(+)}(r) {\sf \hat{A}} - {\sf q}_f^{(-)}(r) \right]
\xi \; {\sf Y}(\vnk) \; .
\nonumber
\end{equation}
If we note that the plane-wave expansion can be written
\begin{equation}
e^{i \vck \cdot \vcr} = 2 \pi i \; {\sf Y}(\vnr)^T
\left[ {\sf q}_f^{(+)}(r) - {\sf q}_f^{(-)}(r) \right]
\xi \; {\sf Y}(\vnk) \; ,
\label{eq:planewavexpqq}
\end{equation}
then it is plain that we may write
\begin{equation}
\chi^+(\vcr) \sim \phi_0(\vcr) +
2 \pi i \; {\sf Y}(\vnr)^T
{\sf q}_f^{(+)}(r) \left[ {\sf \hat{A}} - {\sf I} \right]
\xi \; {\sf Y}(\vnk) \; .
\nonumber
\end{equation}
We now use the asymptotic form of the spherical Hankel function,
\begin{eqnarray}
h^+_l(x) \rightarrow \frac{1}{x} e^{i[x-\frac{1}{2}(l+1)\pi]} &
\mbox{as}  &  r \rightarrow \infty \; ,
\nonumber
\end{eqnarray}
to show that, at large $r$,
\begin{equation}
{\sf q}_f^{(+)}(r) \sim - \frac{1}{k r} e^{i k r} \xi^{\dagger} \; .
\nonumber
\end{equation}
We may then write finally
\begin{equation}
\chi^+(\vcr) \sim \phi_0(\vcr) +
\left( \frac{2 \pi i}{k} \right) \; {\sf Y}(\vnr)^T
\left[ {\sf I} - {\sf A} \right] {\sf Y}(\vnk)
\left\{ \frac{e^{i k r}}{r} \right\} \; ,
\label{eq:chiasymp}
\end{equation}
where we have defined
\begin{equation}
{\sf A} \equiv \xi^{\dagger} {\sf \hat{A}} \xi
= \xi^{\dagger} \; {{\sf M}_-}^T
\left( {\sf M}_+^{-1} \right)^{T} \xi \; ,
\label{eq:scatampmtxa}
\end{equation}
also plainly a unitary matrix.

Having expressed the asymptotic form of $\chi^+$ in this way we
recognize immediately that the usual scattering amplitude is given by
\begin{equation}
f(\vck',\vck) =
(2 \pi i / k) \; {\sf Y}(\vnk')^T
\left[ {\sf I} - {\sf A} \right] {\sf Y}(\vnk) \; ,
\label{eq:fkk}
\end{equation}
where $\vck'$ is of the same magnitude as $\vck$, but whose
direction $\vnr$ is that at which one observes the scattering from
the incoming wave of direction $\vnk$. We note here that this result
agrees essentially with the corresponding expression in the
interesting work of Ziegler~\cite{Ziegle89a}. It can be fairly
easily shown that our matrix ${\sf A}$ is the same,
within some simple matrix factors, as his $\Gamma$.
His development is a direct approach in terms of asymptotic
forms of the solutions of the coupled radial Schr\"{o}dinger
equations, rather than, as here, a straightforward application
of our explicit general Green-function formula. It should be noted,
however, that the most important result of our present development is
not so much the asymptotic form (\ref{eq:chiasymp}) of $\chi^+$,
which is well known and may be derived from more general principles,
but the formula (\ref{eq:scatampmtxa}) for the matrix ${\sf A}$
expressing it in terms of directly calculable quantities.

Of course, it is not always necessary to have an explicit form
for ${\sf A}$ to derive some useful results. It is therefore
interesting to verify the usual well known properties of $f$ directly
from the formula (\ref{eq:fkk}). Thus, to show the
{\it reciprocity theorem} (see \cite{Schiff68a} page 135)
we note that
\begin{eqnarray}
f(-\vck,-\vck') & = & (2 \pi i / k) {\sf Y}(-\vnk)^T
    \left[ {\sf I} - {\sf A} \right] {\sf Y}(-\vnk') \nonumber \\
& = & (2 \pi i / k) {\sf Y}(\vnk)^T \eta
    \left[ {\sf I} - {\sf A} \right] \eta {\sf Y}(\vnk') \nonumber \\
& = & (2 \pi i / k) {\sf Y}(\vnk')^T
    \left[ {\sf I} - \eta {\sf A}^T \eta \right]
    {\sf Y}(\vnk) \nonumber \\
& = & f(\vck',\vck) \; ;
\end{eqnarray}
where, in the first step we used the parity of spherical harmonics,
in the second we took the matrix transpose, and we justify the
final step by observing that
\begin{equation}
\eta {\sf A}^T \eta = \eta \xi {\sf \hat{A}}^T \xi^{\dagger} \eta
= \xi^{\dagger} {\sf \hat{A}} \xi = {\sf A} \; ,
\nonumber
\end{equation}
using the symmetry of ${\sf \hat{A}}$ and the properties of $\xi$
and $\eta$ [Eqs. (\ref{eq:xietadefs}) and (\ref{eq:xietaprops})].

Furthermore we have that
\begin{eqnarray}
\int f(\vck_r,\vck')^* f(\vck_r,\vck) \; d\Omega_{\vck_r} &
= & (4 \pi^2 / k^2) \int {\sf Y}(\vnk')^T \left[ {\sf I}
- {\sf A}^{\dagger} \right] {\sf Y}(\vnk_r)
{\sf Y}(\vnk_r)^T \left[ {\sf I} - {\sf A} \right] {\sf Y}(\vnk)
\; d\Omega_{\vck_r}  \nonumber \\
& = & (4 \pi^2 / k^2) {\sf Y}(\vnk')^T
\left[ {\sf I} - {\sf A}^{\dagger} - {\sf A} + {\sf I} \right] 
{\sf Y}(\vnk)  \nonumber \\
& = & (2 \pi i / k) \left[ f(\vck,\vck')^*
- f(\vck',\vck) \right] \; ;
\end{eqnarray}
where, in the first equality we have matrix transposed the (scalar)
$f^*$ term; for the next step we used spherical-harmonic
orthonormality, multiplied out the matrix terms and used the
unitarity of ${\sf A}$; and in the last step split the
linear sum into two parts. We recognize in this result the
{\it generalized optical theorem}. As is well known, setting $\vck'$
to $\vck$ in this result, and observing that the differential
cross section is given by
\begin{equation}
\sigma(\vck_r,\vck) = \left| f(\vck_r,\vck) \right|^2 ,
\end{equation}
leads directly to an expression for the total elastic scattering cross
section,
\begin{equation}
\sigma_{tot} = \int \sigma(\vck_r,\vck) \; d\Omega_{\vck_r} =
(4 \pi / k) \; \mbox{Im} \, f(\vck,\vck) \; ,
\end{equation}
the famous {\it optical theorem}.


\subsection{Green-function origin translations and re-expansions}

The mathematical elaboration of multiple-scattering theory relies to a large
extent on the translation of expansions of various functions, most notably
the Green function, from one scattering center to another. Since the Green
function used is normally the free Green function this poses no problems as
it is translationally invariant, and its expansions and re-expansions in
partial waves are well known. In our case the problem is more complicated
since our potential need have no translational or rotational symmetry, and
this will be reflected in the Green function associated with it. It is
nevertheless necessary to know how the functional form of the Green function
appears seen from different centers. It is useful in this context therefore
to highlight here a couple of basic mathematical issues that underpin the
justification for some of the steps in our arguments below.

The first point that we need to deal with is uniqueness of the Green
function. It is, of course, well known that the inhomogeneous
Schr\"{o}dinger equation (\ref{eq:gdef}) does not
specify the Green function uniquely, since the boundary conditions
must be imposed. It is easy to show that a solution of Eq.~(\ref{eq:gdef}),
viewed from a translated origin, will also be a solution
of the corresponding equation with the original potential viewed from
the new origin. Intuitively one suspects that, since the boundary
conditions are imposed at $r \rightarrow \infty$, and are the same
in both cases, the resulting solutions will be just different points
of view of one unique Green function;
however, this needs to be shown formally.

In fact it is the Lippmann-Schwinger equation for the Green function,
\begin{equation}
G^+(\vcr,\vcs) = G_0^+(\vcr,\vcs) + \int G_0^+(\vcr,\vct) V(\vct)
G^+(\vct,\vcs) \; d^3\!t \, ,
\label{eq:lippschwgf}
\end{equation}
or equivalently,
\begin{equation}
G^+(\vcr,\vcs) = G_0^+(\vcr,\vcs) + \int G^+(\vcr,\vct) V(\vct)
G_0^+(\vct,\vcs) \; d^3\!t \, ,
\label{eq:lippschwgfalt}
\end{equation}
which guarantees uniqueness (see Ref.~\cite{Taylor72a} page 133 or
for a detailed proof Ref.~\cite{Prugov71a} page 490 ff).

It is not difficult to check that the unique solution $G^+$ of
Eqs. (\ref{eq:lippschwgf}) and (\ref{eq:lippschwgfalt}), when
considered relative to a translated origin of coordinates, satisfies
the corresponding equation in the new coordinate system.
(Of course, the potential $V$ will also have a different functional
form in the new coordinates.) This uniqueness guarantees that when
we find a form for $G^+$ in some coordinate system, it will be equal
to a translated version of the form arising with another origin.

It is useful to check directly that our general form (\ref{eq:gfinal})
does indeed satisfy Eqs. (\ref{eq:lippschwgf}) and
(\ref{eq:lippschwgfalt}). Thus let us assume $G^+$ is of the form of
Eq. (\ref{eq:gfinal}) and consider Eq. (\ref{eq:lippschwgf})
for example. Then we should have
\begin{eqnarray}
G^+(\vcr,\vcs) & = & G_0^+(\vcr,\vcs) + \int G_0^+(\vcr,\vct) V(\vct)
G^+(\vct,\vcs) \; d^3\!t \nonumber \\
& = & \int G_0^+(\vcr,\vct) \left( {\nabla_{\vct}}^2 + E \right)
G^+(\vct,\vcs) \; d^3\!t \nonumber \\
& = & G^+(\vcr,\vcs) +
\int \left[ G_0^+(\vcr,\vct)
\left( {\nabla_{\vct}}^2 + E \right) G^+(\vct,\vcs) -
G^+(\vct,\vcs)
\left( {\nabla_{\vct}}^2 + E \right) G_0^+(\vcr,\vct)
\right] \; d^3\!t \nonumber \\
& = & G^+(\vcr,\vcs) +
\lim_{R \rightarrow \infty} \int \left.
\! \left\{ G_0^+(\vcr,\vct)
\frac{\partial}{\partial t} \left[
G^+(\vct,\vcs) \right] - G^+(\vct,\vcs)
\frac{\partial}{\partial t} \left[
G_0^+(\vcr,\vct) \right] 
\right\} \right|_{t=R}
\! R^2 \, d\Omega_{\vct} ; \nonumber
\end{eqnarray}
where, in the first step we have replaced the $VG^+$ in a similar way
to that of Eq. (\ref{eq:chipls}) [which is justified because our form
for $G^+$ was developed directly from Eq. (\ref{eq:gdef})], in the
second we use a similar property of $G_0^+$, and in the third we invoke
Green's theorem for the region within a sphere of radius $R$ (taken to
be greater than $r$ and $s$) which we allow to go to infinity. Clearly,
for the equation to be true we require that the limit of the integral
be zero. Using the appropriate expansions for the Green functions,
noting that $t=R>r,s$, we may rewrite this (making use also of
spherical-harmonic orthonormality) as the expression
\begin{equation}
{\sf Y}(\vnr)^T
{\sf p}_f(r) \left( {\sf M}_{+f}^{-1} \right)^{T}
\left[ \lim_{R \rightarrow \infty} \left\{
{\sf q}_f^{(+)}(R)^T {{\sf q}^{(+)}}'(R)
- {{\sf q}_f^{(+)}}'(R)^T {\sf q}^{(+)}(R)
\right\} R^2 \right]
{\sf M_+}^{-1} {\sf p}(s)^T {\sf Y}(\vns) \; .
\label{eq:greenfndefncheck}
\end{equation}
Now, since ${\sf q}^{(+)}$ was chosen to be asymptotic to the
free-particle case ${\sf q}_f^{(+)}$ (assuming as usual a potential
which tends quickly enough to zero),
the Wronskian will tend to zero as desired.
So, with the foregoing arguments in mind, we may have confidence
that the form of $G^+$ that we derive in some local coordinate system
is identical to that found in some translated coordinates.


\subsubsection{Re-expansion of free Green functions}

Let us now derive the re-expansion formulas for the free Green function
$G_0^+$. These are, of course, well known (and may be found in
Ref.~\cite{Johnso73a} whose approach we follow closely). However, it is
useful to repeat briefly the derivation, since it allows us to introduce
some new notation and express the formulas in a different form than usual.
Also it is a model for our later derivations.

First, we recall the atomic centers $\vcR_i$ (for $i=1,\ldots,N$) which we
introduced earlier. For some general position vector $\vcr$ let us
denote the relative position vector with respect some given atomic
center $\vcR_i$ by
\begin{equation}
\vcr_i \equiv \vcr - \vcR_i \; .
\end{equation}
For the purpose of developing the re-expansion formulas we may consider,
say, two distinct centers, $\vcR_i$ and $\vcR_j$, as being two general
points in space. The resulting formulas will then be immediately useful
for our multiple-scattering problem.

We consider first the re-expansion of the free radial
Schr\"{o}dinger-equation solutions, ${\sf p}_f(r) \equiv {\sf j}(kr)$,
and begin by considering a plane wave expressed in terms of
coordinates relative to $\vcR_i$ and $\vcR_j$ thus:
\begin{equation}
\exp ( i \vck \cdot \vcr ) =
\exp \left[ i \vck \cdot (\vcr_i + \vcR_i)
\right] = \exp ( i \vck \cdot \vcr_i ) \exp ( i \vck \cdot \vcR_i )  
\nonumber
\end{equation}
and
\begin{equation}
\exp ( i \vck \cdot \vcr ) = \exp ( i \vck \cdot \vcr_j )
\exp ( i \vck \cdot \vcR_j ) \; .
\nonumber
\end{equation}
Equating the two expressions and using the plane-wave expansion
(\ref{eq:planewavexp}) leads, with some rearrangement, to
\begin{equation}
{\sf Y}(\vnr_i)^T {\sf j}(kr_i) \, \xi \, {\sf Y}(\vnk) =
{\sf Y}(\vnr_j)^T {\sf j}(kr_j) \, \xi \, {\sf Y}(\vnk)
\exp ( i \vck \cdot \vcR_{ij} ) \, ,
\nonumber
\end{equation}
where we have defined the atomic center-to-center vector by
\begin{equation}
\vcR_{ij} \equiv \vcR_{j} - \vcR_{i} \; .
\end{equation}
If we multiply both sides of our equation on the right by
${\sf Y}(\vnk)^T$ and integrate over all $\vnk$, then we get
\begin{equation}
{\sf Y}(\vnr_i)^T {\sf j}(kr_i) =
{\sf Y}(\vnr_j)^T {\sf j}(kr_j) \, {\sf D}(k;\vcR_{ij}) \, ,
\label{eq:reexpnp0}
\end{equation}
where we have used spherical-harmonic orthonormality and defined the
matrix associated with displacement, for general $\vcR$, by
\begin{equation}
{\sf D}(k;\vcR) \equiv \xi \, {\sf \hat{D}}(k;\vcR)
\, \xi^{\dagger}
\end{equation}
and
\begin{equation}
{\sf \hat{D}}(k;\vcR) \equiv
\int {\sf Y}(\vnk)
\exp ( i \vck \cdot \vcR ) 
{\sf Y}(\vnk)^T \; d\Omega_{\vck} \; .
\label{eq:dhatdef}
\end{equation}
If we use the plane-wave expansion for the exponential in the last
equation, evaluate the angular integrals, and rearrange, we find that
the matrix elements of ${\sf \hat{D}}(k;\vcR)$ are given by
\begin{equation}
\hat{D}_{L L'}(k;\vcR) = 4 \pi \sum_{L''} i^{l''} j_{l''}(kR)
Y_{L''}(\vnR) I(L,L',L'') \; ,
\label{eq:dhatmxel}
\end{equation}
from which one may note that ${\sf \hat{D}}(k;\vcR)$ is symmetric.
The matrix elements of ${\sf D}(k;\vcR)$ are now easily seen to be
\begin{equation}
D_{L L'}(k;\vcR) = 4 \pi \sum_{L''} i^{l''-l'+l} j_{l''}(kR)
Y_{L''}(\vnR) I(L,L',L'') \; .
\label{eq:dmxel}
\end{equation}
If we now note that the Gaunt coefficients $I(L,L',L'')$ are non-zero
only when $l+l'+l''$ is an even integer, then we see that
${\sf D}(k;\vcR)$ is real. Writing out fully the summations in
Eq. (\ref{eq:reexpnp0}), and using the explicit matrix elements
(\ref{eq:dmxel}), allows us to recover the usual re-expansion
formula \cite{Johnso73a}.

Now since the points indexed $i$ and $j$ are arbitrary we may
interchange them to obtain another equation like
Eq. (\ref{eq:reexpnp0}), which we use to eliminate the term
${\sf Y}(\vnr_j)^T {\sf j}(kr_j)$. On rearrangement we then find
\begin{equation}
{\sf Y}(\vnr_i)^T {\sf j}(kr_i) \left[ {\sf I} -
{\sf D}(k;-\vcR_{ij}){\sf D}(k;\vcR_{ij}) \right] = 0 \; .
\nonumber
\end{equation}
Multiplying on the left by ${\sf Y}(\vnr_i)$, integrating over all
angles, and invoking spherical-harmonic orthonormality, then leads to
\begin{equation}
{\sf j}(kr_i) \left[ {\sf I} -
{\sf D}(k;-\vcR_{ij}){\sf D}(k;\vcR_{ij}) \right] = 0 \; .
\nonumber
\end{equation}
Since the diagonal matrix ${\sf j}(kr_i)$ consists of the linear
independent (regular) solutions of the radial Schr\"{o}dinger equation,
then the expression in square brackets must be the zero matrix. In other
words, since $\vcR_{ij}$ is arbitrary, then we have that, in general,
\begin{equation}
{\sf D}(k;\vcR)^{-1} = {\sf D}(k;-\vcR) \; .
\label{eq:dkrinv}
\end{equation}
Furthermore, we have
\begin{eqnarray}
{\sf D}(k;\vcR)^{-1}
& = & \xi \, {\sf \hat{D}}(k;-\vcR) \, \xi^{\dagger} \nonumber \\
& = & \xi \, {\sf \hat{D}}(k;\vcR)^* \, \xi^{\dagger} \nonumber \\
& = & \xi \, {\sf \hat{D}}(k;\vcR)^{\dagger}
      \, \xi^{\dagger} \nonumber \\
& = & {\sf D}(k;\vcR)^{\dagger} \nonumber \; ,
\end{eqnarray}
where, in the first step we take the complex conjugate of
Eq.~(\ref{eq:dhatdef}), and in the second we invoked the symmetry of
${\sf \hat{D}}(k;\vcR)$. Thus ${\sf D}(k;\vcR)$ is unitary.
Since it is real, then it is orthogonal. It may be also
easily seen that
\begin{equation}
{\sf \hat{D}}(k;\vcR)^{-1} = {\sf \hat{D}}(k;-\vcR)
= {\sf \hat{D}}(k;\vcR)^{\dagger} \; . 
\end{equation}

To develop the corresponding formula for the free (outgoing-wave) radial
solutions regular at infinity, ${\sf q}_f^{(+)}(r) = -i {\sf h}^+(kr)$,
we need to use the translation properties of $G_0^+$. However, since
these solutions are singular at the origin, the re-expansion must be
approached differently above and below the separation distance, which
regions we shall denote as the ``far'' and ``near'' regions,
respectively.

\begin{figure}
\includegraphics[scale=0.800]{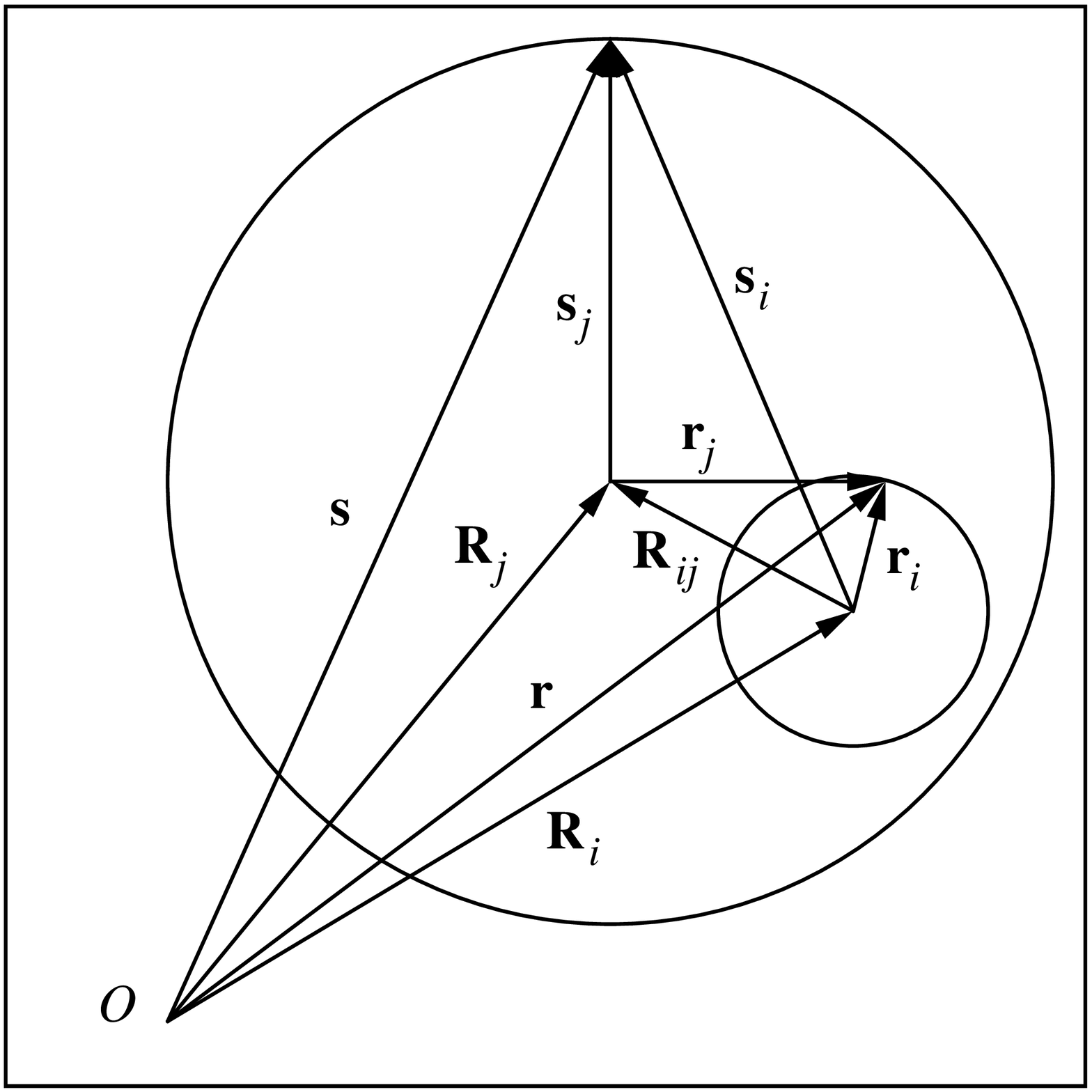} 
\caption{Diagram of vectors and sphere surfaces in the
derivation of the re-expansion of the irregular radial solutions
in the far region.}
\label{fig-1}
\end{figure}

With reference to the diagram in Fig.~\ref{fig-1}, let us consider
a general point with position vector $\vcs$, on the surface of a sphere
centered on $\vcR_j$, with a radius somewhat greater than $R_{ij}$.
Let $\vcr$ represent a general point on a sphere of
positive radius centered on $\vcR_i$, contained
entirely within the first sphere. Such a sphere
can always be found, since $s_j > R_{ij}$. It may be then seen that
in this configuration we shall always have $s_i > r_i$ and $s_j > r_j$.
Therefore we may write
\begin{equation}
G^+_0(\vcr,\vcs) = - \frac{e^{i k  \left| \vcr_i - \vcs_i \right|}}
{4 \pi \left| \vcr_i - \vcs_i \right|} =
-i k {\sf Y}(\vnr_i)^T {\sf j}(k r_i)
{\sf h}^+(k s_i) {\sf Y}(\vns_i) \; .
\nonumber
\end{equation}
Translation invariance allows to write also, since
$\vcr_i-\vcs_i = \vcr_j-\vcs_j$, that
\begin{equation}
G^+_0(\vcr,\vcs) = - \frac{e^{i k  \left| \vcr_j - \vcs_j \right|}}
{4 \pi \left| \vcr_j - \vcs_j \right|} =
-i k {\sf Y}(\vnr_j)^T {\sf j}(k r_j)
{\sf h}^+(k s_j) {\sf Y}(\vns_j) \; .
\nonumber
\end{equation}
Equating the two right-hand expressions gives us
\begin{equation}
{\sf Y}(\vnr_i)^T {\sf j}(k r_i) {\sf h}^+(k s_i) {\sf Y}(\vns_i) =
{\sf Y}(\vnr_j)^T {\sf j}(k r_j) {\sf h}^+(k s_j) {\sf Y}(\vns_j) \; ,
\nonumber
\end{equation}
which may be rewritten, using Eq. (\ref{eq:reexpnp0}), as
\begin{equation}
{\sf Y}(\vnr_i)^T {\sf j}(k r_i)
\left[ {\sf h}^+(k s_i) {\sf Y}(\vns_i)
 - {\sf D}(k;-\vcR_{ij}) {\sf h}^+(k s_j) {\sf Y}(\vns_j)
\right] = 0 \; .
\nonumber
\end{equation}
Since the sphere associated with $\vcr$ is entirely within that
associated with $\vcs$, then the last equation is true for all
$\vnr_i$ and at least a small range of $r_i$. So we may invoke
our previous argument to conclude that the expression in square
brackets (which has no dependence on $\vcr_i$) is zero.
Thus, transposing and using the properties of ${\sf D}(k;\vcR)$,
we may write
\begin{eqnarray}
{\sf Y}(\vns_i)^T {\sf h}^+(k s_i) =
{\sf Y}(\vns_j)^T {\sf h}^+(k s_j) {\sf D}(k;\vcR_{ij})
& \mbox{for} & s_j > R_{ij} \; ,
\label{eq:reexpq0far}
\end{eqnarray}
the far-region re-expansion formula. This may be compared with
Eq. (\ref{eq:reexpnp0}) (indeed the real part corresponds identically).

To obtain the re-expansion formula for the near region we first write
out Eq. (\ref{eq:reexpq0far}) explicitly, so that we have
\begin{equation}
h^+_l(k s_i) Y_L(\vns_i) =
4 \pi \sum_{L' L''} i^{l''-l+l'} j_{l''}(kR_{i j})
h^+_{l'}(k s_j) 
Y_{L''}(\vnR_{i j}) 
Y_{L'}(\vns_j) 
I(L,L',L'') \; ;
\end{equation}
where $\vcs_i = \vcs_j + \vcR_{i j}$ with the restriction $s_j > R_{ij}$. In
the near-region case, where $s_j < R_{ij}$, we may use the same formula with
the variables interchanged, since the vectors are arbitrary apart from the
order relation. Therefore, we have
\begin{equation}
h^+_l(k s_i) Y_L(\vns_i) =
4 \pi \sum_{L' L''} i^{l''-l+l'} 
j_{l''}(k s_j) 
h^+_{l'}(kR_{i j})
Y_{L''}(\vns_j) 
Y_{L'}(\vnR_{i j}) I(L,L',L'') \; .
\end{equation}
This may be written in our notation as
\begin{eqnarray}
{\sf Y}(\vns_i)^T {\sf h}^+(k s_i) =
{\sf Y}(\vns_j)^T {\sf j}(k s_j) {\sf F}(k;\vcR_{ij})
& \mbox{for} & s_j < R_{ij} \; ,
\label{eq:reexpq0near}
\end{eqnarray}
where we have defined ${\sf F}(k;\vcR_{ij})$ by
\begin{equation}
F_{L L'}(k;\vcR_{ij}) \equiv 
4 \pi \sum_{L'} i^{l-l'+l''} 
h^+_{l''}(kR_{i j})
Y_{L''}(\vnR_{i j}) I(L,L',L'') \; .
\end{equation}


\subsubsection{Gaunt algebra}

Having encountered the Gaunt coefficients yet again in our formulas
it seems worthwhile to investigate them in more detail at this point.
We shall look briefly at some of their properties, but consider
them from a slightly unusual point of view.

Normally the Gaunt coefficients arise in the re-expansion of the
product of two spherical harmonics, where
\begin{equation}
Y_L(\vnr) Y_{L'}(\vnr) = \sum_{L''} I(L,L',L'') Y_{L''}(\vnr) \; .
\label{eq:shprodexp}
\end{equation}
Use of spherical-harmonic orthonormality leads immediately to the
definition (\ref{eq:gauntdef}) that we presented earlier. In the case of
complex spherical harmonics the corresponding versions of the coefficients
have a well known expression in terms of Clebsch-Gordan coefficients.

The particular form of the definitions of
matrices such as ${\sf w}$ of Eq. (\ref{eq:potmx}),
${\sf \hat{D}}(k;\vcR)$ of Eq. (\ref{eq:dhatdef}), 
and ${\sf F}(k;\vcR)$ of Eq. (\ref{eq:reexpq0near}), for example,
suggests that it may be useful to consider a set of ``Gaunt matrices''
${\sf \Gamma}^{(L)}$, indexed by $L$, with matrix elements defined by
\begin{equation}
{\Gamma^{(L'')}}_{L L'} \equiv I(L,L',L'') \; ,
\end{equation}
easily seen to be real and symmetric.
We may now reinterpret Eq. (\ref{eq:shprodexp}) as an eigenvalue equation,
\begin{equation}
[Y_L(\vnr)] {\sf Y}(\vnr) = {\sf \Gamma}^{(L)} {\sf Y}(\vnr) \; ,
\label{eq:shprodexpeig}
\end{equation}
in which, for any $\vnr$, the vector ${\sf Y}(\vnr)$ is an eigenvector
of the matrix ${\sf \Gamma}^{(L)}$, with eigenvalue $Y_L(\vnr)$.

Now the properties of the Gaunt coefficients allow us to say something
about the structure of the individual matrices. In particular,
we note that the triangle condition on $l$, $l'$, and $l''$ implies
that the $L$th row (column) of ${\sf \Gamma}^{(L'')}$ can only have
a finite number of nonzero entries, from index $l' = |l'' - l|$ up to
index $l' = l'' + l$. Furthermore we note that the matrix is banded
with respect to $l$.

Of special note is the fact that the matrix elements of the first row
(column) are given by 
\begin{equation}
{\Gamma^{(L'')}}_{(0,0) L'} = I\left((0,0),L',L''\right)
= (4 \pi)^{-1/2} \delta_{L' L''}\; ,
\label{eq:gammalrow1}
\end{equation}
from which we may deduce
that the Gaunt matrices are linearly independent.

For a given vector ${\sf a}$ with (in general complex) components $a_L$,
let us define a linear combination of the Gaunt matrices by
\begin{equation}
{\sf \Lambda}[{\sf a}] \equiv \sum_L a_L {\sf \Gamma}^{(L)} \; .
\label{eq:gauntlincom}
\end{equation}
There is no need to consider convergence in this definition since each
matrix element of ${\sf \Lambda}[{\sf a}]$ only contains contributions
from a finite number of the $a_L$, by virtue of what was said above.
Linear independence gives us that
\begin{equation}
{\sf \Lambda}[{\sf a}] = 0 \Leftrightarrow {\sf a} = 0 \; .
\end{equation}
If one is given a ${\sf \Lambda}[{\sf a}]$, one may recover ${\sf a}$ by
reading off the first row (column) and using Eq. (\ref{eq:gammalrow1});
which shows that the correspondence between ${\sf a}$ and
${\sf \Lambda}[{\sf a}]$ is one-to-one.
It is also not difficult to show that ${\sf \Lambda}[{\sf a}]$ is
a linear function of its parameter ${\sf a}$, and that, if the products
exist and are finite,
\begin{equation}
{\sf \Lambda}[{\sf a}] {\sf b} = {\sf \Lambda}[{\sf b}] {\sf a} \; .
\end{equation}
Among the cases that we have encountered we may write, for example,
\begin{equation}
{\sf \hat{D}}(k;\vcR) = {\sf \Lambda}[4 \pi \xi {\sf j}(k R)
{\sf Y}(\vnR)] \; .
\label{eq:dhatlambda}
\end{equation}

To see some more interesting properties let us consider now the
integral of the product of four spherical harmonics,
\begin{equation}
I_4(L,L',L'',L''') \equiv \int Y_L(\vnt) Y_{L'}(\vnt)
Y_{L''}(\vnt) Y_{L'''}(\vnt) \; 
d\Omega_{\vct} \; ,
\end{equation}
which are manifestly symmetric under all permutations of the indices.
We may use Eq. (\ref{eq:shprodexp}) to substitute for pairs of
spherical harmonics in this definition, then use orthonormality
to obtain three possible equivalent expressions, so that
\begin{eqnarray}
I_4(L,L',L'',L''')
& = & \sum_{L_1} I(L,L''',L_1) I(L',L'',L_1) \nonumber \\
& = & \sum_{L_1} I(L,L'',L_1)  I(L''',L',L_1) \nonumber \\
& = & \sum_{L_1} I(L,L',L_1)   I(L'',L''',L_1) \; .
\label{eq:gmxprodelts}
\end{eqnarray}
Making a particular choice for the assignment of indices leads to the
matrix equations
\begin{equation}
{\sf \Gamma}^{(L)} {\sf \Gamma}^{(L')}
= {\sf \Gamma}^{(L')} {\sf \Gamma}^{(L)}
= \sum_{L_1} I(L,L',L_1) {\sf \Gamma}^{(L_1)} \; .
\label{eq:gauntprops}
\end{equation}
In summary then, the ${\sf \Gamma}^{(L)}$ form a set of real,
symmetric, linearly independent,
mutually commuting matrices, the product of any pair of which may be
expressed as a linear combination of some of them (in fact in the same
way as the spherical harmonics from which they derive). The fact that
they commute with each other is, of course, consistent with their
having common eigenvectors.

It is interesting now to consider the ${\sf \Gamma}^{(L)}$ as
generators of some larger set of matrices. In particular, let us denote
by $A_G$ the set of all linear combinations ${\sf \Lambda}[{\sf a}]$,
i.e., the set spanned by the ${\sf \Gamma}^{(L)}$. By virtue of the
relations (\ref{eq:gauntprops}) we see that this set is closed under
matrix multiplication; although, when we consider the product of two
elements of $A_G$, we must be careful here about convergence.
Indeed, we see that the general matrix element of a product of
linear combinations may be written
\begin{equation}
\left( {\sf \Lambda}[{\sf a}] {\sf \Lambda}[{\sf b}] \right)_{L L'}
= \int Y_L(\vnt)
\left[ \sum_{L''} a_{L''} Y_{L''}(\vnt) \right]
\left[ \sum_{L'''} b_{L'''} Y_{L'''}(\vnt) \right]
Y_{L'}(\vnt) \; d\Omega_{\vct} \; ,
\nonumber
\end{equation}
which, to be finite, requires that the admissible vectors for $A_G$
represent convergent spherical harmonic expansions of, say,
square-integrable functions of $\vnr$. Clearly $A_G$ is a commutative
algebra of matrices which we shall call the ``Gaunt algebra''. (We
conjecture also that is possible that the $l=1$ Gaunt matrices
generate all the rest.)

It is not difficult to show, using the equations (\ref{eq:gmxprodelts}),
that we have
\begin{equation}
{\sf \Lambda}[{\sf \Lambda}[{\sf a}] {\sf b}]
= {\sf \Lambda}[{\sf a}] {\sf \Lambda}[{\sf b}] \; .
\end{equation}
This may be used to show a further property of ${\sf \hat{D}}(k;\vcR)$.
To see this rewrite the re-expansion formula (\ref{eq:reexpnp0}) as
\begin{equation}
\xi \, {\sf j}(kr_i) {\sf Y}(\vnr_i) =
{\sf \hat{D}}(k;\vcR_{ij}) \, \xi \, {\sf j}(kr_j) {\sf Y}(\vnr_j) \; .
\nonumber
\end{equation}
Using Eq. (\ref{eq:dhatlambda}) we see straightforwardly that
\begin{eqnarray}
{\sf \hat{D}}(k;\vcr_{i}) & = & {\sf \Lambda}[4 \pi \xi {\sf j}(k r_i)
{\sf Y}(\vnr_i)] \nonumber \\
& = & {\sf \Lambda}[{\sf \hat{D}}(k;\vcR_{ij}) 4 \pi \xi {\sf j}(kr_j)
{\sf Y}(\vnr_j) ] \nonumber \\
& = & {\sf \hat{D}}(k;\vcR_{ij}) {\sf \Lambda}[4 \pi \xi {\sf j}(kr_j)
{\sf Y}(\vnr_j) ] \nonumber \\
& = & {\sf \hat{D}}(k;\vcR_{ij}) {\sf \hat{D}}(k;\vcr_{j}) \; .
\nonumber
\end{eqnarray}
Now $\vcr_i = \vcR_{i j} + \vcr_j$ with no restrictions, and the vectors
are effectively arbitrary. So we see that, for general vectors $\vcR$
and $\vcS$, we have 
\begin{equation}
{\sf \hat{D}}(k;\vcR+\vcS) = {\sf \hat{D}}(k;\vcR)
{\sf \hat{D}}(k;\vcS) \; ,
\label{eq:sumprod}
\end{equation}
and it is easy to see that the corresponding relation holds for
${\sf D}(k;\vcR)$. We note here that this relation can be proved directly
from the definition (\ref{eq:dhatdef}) if one makes use of the angular
delta-function property of spherical harmonics,
\begin{equation}
\sum_L Y_L(\vnr) Y_L(\vns) = {\sf Y}(\vnr)^T {\sf Y}(\vns)
= \delta(\vnr - \vns) \; .
\label{eq:angdelta}
\end{equation}
This property has been noted before by Natoli {\it et al.}
in Ref. \cite{Na+al.90a} [see their Appendix A noting that our
$D_{L L'}(k;\vcR_{i j})$ corresponds to $J_{L L'}^{i j}$
in the notation of their Eq.~(A23)].


\subsubsection{Re-expansion of general regular solutions}

We may now derive the re-expansions of the general coupled radial
Schr\"{o}dinger-equation solutions ${\sf p}$ and ${\sf q}$.
For the general regular solutions ${\sf p}$ we consider the
explicit form (\ref{eq:lippschwdwfnex}) of the Lippmann-Schwinger 
equation and the resulting scattering state $\chi^+(\vcr)$
of Eq.~(\ref{eq:chip}), noting that the plane-wave term $\phi_0(\vcr)$
occurs linearly on the rhs of the former. If we now consider an initial
plane wave $\phi_{i 0}(\vcr)$ given by
\begin{equation}
\phi_{i 0}(\vcr) \equiv e^{i \vck \cdot \vcr_i}
= e^{i \vck \cdot \vcr} e^{- i \vck \cdot \vcR_i}
= e^{- i \vck \cdot \vcR_i} \phi_{0}(\vcr) \; ,
\end{equation}
then we may repeat the derivation of Eq.~(\ref{eq:chip}) with respect
to a new origin at $\vcR_i$. This leads to a scattering state
\begin{equation}
\chi^+_i(\vcr_i) =
( {4 \pi}/{k}) \, {\sf Y}(\vnr_i)^T {\sf p}_i(r_i)
\left( {\sf M}_{i +}^{-1} \right)^{T}
\xi \, {\sf Y}(\vnk) \; ,
\label{eq:chipi}
\end{equation}
where the subscript $i$ on ${\sf p}_i$, ${\sf M}_{i +}$ (and implicitly
${\sf q}_i^{(+)}$) shows that the coupled radial Schr\"{o}dinger-equation
solutions with respect to the center $\vcR_i$ will be, in general,
different to those of other centers.
If we now invoke the linearity with respect to $\phi_0(\vcr)$ of the rhs
of Eq.~(\ref{eq:lippschwdwfnex}), and the uniqueness of solutions of the
Lippmann-Schwinger equation (see for example Ref.~\cite{Prugov71a}),
then we may conclude that
\begin{equation}
\chi^+(\vcr) = e^{i \vck \cdot \vcR_i} \chi^+_i(\vcr_i) \; .
\nonumber
\end{equation}
(Since the initial plane waves differ only by a constant phase factor, it
should be intuitively obvious that the scattered waves must be essentially
the same within a phase factor.) Therefore we have
\begin{equation}
\chi^+(\vcr) =
( {4 \pi}/{k}) \, {\sf Y}(\vnr_i)^T {\sf p}_i(r_i)
\left( {\sf M}_{i +}^{-1} \right)^{T}
\xi \, {\sf Y}(\vnk) \: e^{i \vck \cdot \vcR_i}  \; .
\nonumber
\end{equation}
A similar result may be derived with respect to some other center
$\vcR_j$, and we can equate the two expressions to obtain
\begin{equation}
{\sf Y}(\vnr_i)^T {\sf p}_i(r_i)
\left( {\sf M}_{i +}^{-1} \right)^{T}
\xi \, {\sf Y}(\vnk) \: e^{i \vck \cdot \vcR_i} =
{\sf Y}(\vnr_j)^T {\sf p}_j(r_j)
\left( {\sf M}_{j +}^{-1} \right)^{T}
\xi \, {\sf Y}(\vnk) \: e^{i \vck \cdot \vcR_j}  \; .
\nonumber
\end{equation}
From this point we may proceed as in the free Green-function case
and find that
\begin{equation}
{\sf Y}(\vnr_i)^T {\sf p}_i(r_i)
\left( {\sf M}_{i +}^{-1} \right)^{T} =
{\sf Y}(\vnr_j)^T {\sf p}_j(r_j)
\left( {\sf M}_{j +}^{-1} \right)^{T}
{\sf D}(k;\vcR_{ij}) \, ,
\label{eq:reexpnpij}
\end{equation}
with no restrictions on the vectors involved. This is then the re-expansion
formula for the general regular solutions. It is easy to see that it reduces
immediately to the result (\ref{eq:reexpnp0}) when then scattering potential
vanishes.

We make two observations about our formula (\ref{eq:reexpnpij}). First,
it may be noted that the radial solution matrices ${\sf p}$ are required
only to be linearly independent, real, and regular at the origin. We do not
specify any further boundary conditions or particular form at $r=0$.
At first sight this may be thought to pose a problem when we relate
solutions centered on different origins whose boundary conditions may be
arbitrarily chosen. In fact this problem is avoided because of the
presence of the inverse of the constant Wronskian matrix ${\sf M}$
for each center, which serves to ``standardize'' the regular solutions
by reference, through the ${\sf q}^{(+)}$, to the common boundary
conditions at infinity.

Second, we note for the free radial functions that the re-expansion
formula has a certain symmetry because both the functions themselves
and the displacement operator ${\sf D}(k;\vcR_{ij})$ involve the
spherical Bessel functions. In fact this symmetry underlies the
product formula (\ref{eq:sumprod}). As a consequence the expressions
on both sides of the formula may be easily
seen to satisfy the free Schr\"{o}dinger equation when considered as
functions of $\vcr_i$, $\vcr_j$ or $\vcR_{i j}$.
If one wishes to investigate the analogous properties of our latest
expression, then one must take into account the ``hidden'' dependence
of ${\sf p}_i$ on $\vcR_i$.


\subsubsection{Re-expansion of general irregular solutions in the far region}

With the result of the previous section we may proceed quickly to
obtain the re-expansion formula for general irregular solutions in the
far region. To see this we consider again Fig.~\ref{fig-1},
where the position vector $\vcs$ is such that
$s_j > R_{i j}$, and we have $s_i > r_i$ and $s_j > r_j$. Expressing the
general Green function in terms of vectors relative to the centers $\vcR_i$
and $\vcR_j$ in turn, using Eq.~(\ref{eq:gfinal}), leads to
\begin{eqnarray}
G^+(\vcr,\vcs) & = &
{\sf Y}(\vnr_i)^T {\sf p}_i(r_i)
\left( {\sf M}_{i +}^{-1} \right)^{T}
{\sf q}^{(+)}_i(s_i)^T {\sf Y}(\vns_i) \nonumber \\
& = &
{\sf Y}(\vnr_j)^T {\sf p}_j(r_j)
\left( {\sf M}_{j +}^{-1} \right)^{T}
{\sf q}^{(+)}_j(s_j)^T {\sf Y}(\vns_j)  \; .
\label{eq:gfnqfar}
\end{eqnarray}
Making use of the previous section's results leads immediately to
\begin{equation}
{\sf Y}(\vnr_i)^T {\sf p}_i(r_i)
\left( {\sf M}_{i +}^{-1} \right)^{T} \left[
{\sf q}^{(+)}_i(s_i)^T {\sf Y}(\vns_i)
- {\sf D}(k;-\vcR_{ij})
{\sf q}^{(+)}_j(s_j)^T {\sf Y}(\vns_j) \right]= 0  \; .
\nonumber
\end{equation}
By similar reasoning to that which led to Eq.~(\ref{eq:reexpq0far}),
we see that the expression inside the square brackets in the last
equation must be zero, and we may then write (on transposing)
\begin{eqnarray}
{\sf Y}(\vns_i)^T {\sf q}^{(+)}_i(s_i) =
{\sf Y}(\vns_j)^T {\sf q}^{(+)}_j(s_j) {\sf D}(k;\vcR_{ij})
& \mbox{for} & s_j > R_{ij} \; ,
\label{eq:reexpqfar}
\end{eqnarray}
the general far-region re-expansion formula. It is also easily seen
to reduce to the free result when the potential vanishes. Taking the
complex conjugate of (\ref{eq:reexpqfar}), noting the reality of the
spherical harmonics and the ${\sf D}(k;\vcR_{ij})$, leads to the
corresponding formula for the incoming-wave solutions:
\begin{eqnarray}
{\sf Y}(\vns_i)^T {\sf q}^{(-)}_i(s_i) =
{\sf Y}(\vns_j)^T {\sf q}^{(-)}_j(s_j) {\sf D}(k;\vcR_{ij})
& \mbox{for} & s_j > R_{ij} \; .
\label{eq:reexpqiwfar}
\end{eqnarray}


\subsubsection{Translation of the single-scattering amplitude}

Our re-expansion formulas for the general irregular solutions in the far
region allow us to take the development of the single-scattering problem
further. In particular we may now elucidate the behavior of the scattering
amplitude as seen with respect to different origins. Intuitively, since
the scattering is calculated from the asymptotic behavior of the solutions,
one expects that a finite displacement should have no real effect. This is
not quite the case as we shall see.

First, however, we must use the formula (\ref{eq:pqpmmqmmp}) to rewrite the
general Green function (\ref{eq:gfinal}) in terms of only the irregular
solutions. With some algebra we find, noting (\ref{eq:defofahat}) and
(\ref{eq:mmmpsym}), that
\begin{eqnarray}
\lefteqn{G(\vcr,\vcs) = } \nonumber \\
 & & \frac{i k}{2} {\sf Y}(\vnr)^T \left[
{\sf q}^{(+)}(r) {\sf \hat{A}} {\sf q}^{(+)}(s)^T 
- {\sf q}^{(-)}(r) {\sf q}^{(+)}(s)^T \theta(s-r)
- {\sf q}^{(+)}(r) {\sf q}^{(-)}(s)^T \theta(r-s)
\right] {\sf Y}(\vns) \; . \nonumber \\
\label{eq:gfinalirreg}
\end{eqnarray}
We now consider this form expressed relative to the centers
$\vcR_i$ and $\vcR_j$ in turn, as before, to give us two equivalent
forms for $G(\vcr,\vcs)$. Let us then take the general vectors $\vcr$
and $\vcs$ to be such that $r_j > R_{i j}$ and $s_j > R_{i j}$,
and use the re-expansion formulas (\ref{eq:reexpqfar}) and
(\ref{eq:reexpqiwfar}) to replace the ${\sf q}^{(+)}_i$ and
${\sf q}^{(-)}_i$ terms in the first of these equations. 
We should also choose $\vcr$ and $\vcs$ to be on spheres centered on
$\vcR_j$ such that, say, $r_j < s_j$ by a margin large enough so that
$r_i < s_i$ as well. This ensures that the theta functions correspond.
Equating the right-hand sides, canceling the theta-function terms,
and rearranging, we find that
\begin{equation}
{\sf Y}(\vnr_j)^T 
{\sf q}^{(+)}_j(r_j) 
\left[ 
{\sf D}(k;\vcR_{ij}) {\sf \hat{A}}_i {\sf D}(k;-\vcR_{ij})
- {\sf \hat{A}}_j 
\right] 
{\sf q}^{(+)}_j(s_j)^T 
{\sf Y}(\vns_j) = 0 \; .
\nonumber
\end{equation}
For large enough $r_j$ and $s_j$ their order relations above are true
for all values of the angular variables, so we can conclude that the
expression in square brackets is zero. Using our earlier definitions
we then find that
\begin{equation}
{\sf \hat{D}}(k;\vcR_{ij}) {\sf A}_i {\sf \hat{D}}(k;-\vcR_{ij})
= {\sf A}_j  \; .
\end{equation}
We have seen in Eq.~(\ref{eq:dhatlambda}) that
${\sf \hat{D}}(k;\vcR_{ij})$ is a linear combination
of Gaunt matrices and as such shares the common eigenvectors
${\sf Y}(\vnr)$ for arbitrary $\vnr$. It is not difficult to
see that the associated eigenvalue is
$\exp(i k \vnr \cdot \vcR_{i j})$, and so
\begin{eqnarray}
{\sf Y}(\vnk')^T {\sf A}_j {\sf Y}(\vnk) & = &
{\sf Y}(\vnk')^T  
{\sf \hat{D}}(k;\vcR_{ij}) {\sf A}_i {\sf \hat{D}}(k;-\vcR_{ij})
{\sf Y}(\vnk) \nonumber \\
& = & \exp[i (\vck' - \vck) \cdot \vcR_{i j}] \;
{\sf Y}(\vnk')^T {\sf A}_i {\sf Y}(\vnk)
\end{eqnarray}
If we now consider the term from Eq.~(\ref{eq:fkk}) for the scattering
amplitude, involving the identity matrix, we see that it is precisely
of the form of Eq.~(\ref{eq:angdelta}), i.e., a delta function of the
angular variables. Therefore it is unchanged if we multiply by
$\exp[i (\vck' - \vck) \cdot \vcR_{i j}]$, since this is a function of
the difference of the same variables and is unity when they coincide.
So finally we may write
\begin{equation}
f_j(\vck',\vck) = \exp[i (\vck' - \vck) \cdot \vcR_{i j}]
f_i(\vck',\vck)  \; .
\label{eq:fkkdisplaced}
\end{equation}
We see from this that the cross section is unchanged, but there is a
phase shift when the scattering amplitude is calculated with respect
to a different origin. The first is entirely in accord with our
expectations, while the second is what gives rise to interference
effects if we had several disjoint scattering potentials.


\subsubsection{Re-expansion of general irregular solutions in the near region}

In the case of the near-region re-expansion formula, we do not yet have a
generally applicable method leading to a closed formula similar to those
of the immediately preceding sections. The argument leading to the
re-expansion formula (\ref{eq:reexpq0near}) for the free Green function
is not available to us in the case of a general potential, since it makes
use of the fact that the free radial solutions are independent of which
center they are referred to.

\begin{figure}
\includegraphics[scale=0.800]{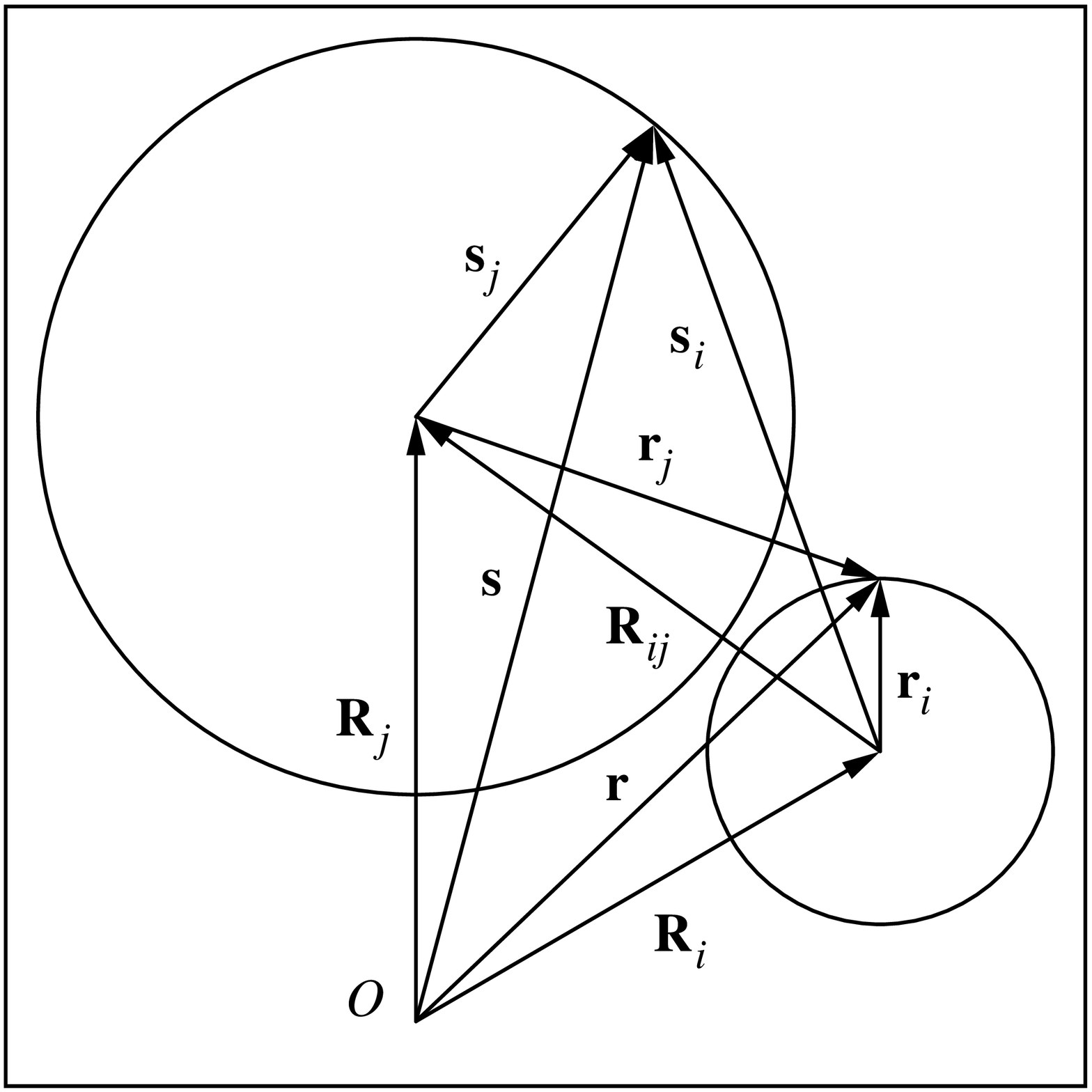} 
\caption{Diagram of vectors and sphere surfaces in the derivation of the
re-expansion of the irregular radial solutions in the near region.}
\label{fig-2}
\end{figure}

Let us therefore look for an alternative approach by
considering the situation in Fig.~\ref{fig-2}, returning to 
the free Green function in the first instance. Thus we choose an
arbitrary general point $\vcs$ on some sphere centered on $\vcR_j$
with, now, $s_j < R_{i j}$; and an arbitrary $\vcr$ on a sphere
centered on $\vcR_i$, entirely outside the first sphere.
Since here we shall always have $s_i > r_i$
and $s_j < r_j$, an argument similar to that preceding Eq.
(\ref{eq:reexpq0far}) leads instead to
\begin{equation}
{\sf Y}(\vnr_i)^T {\sf j}(k r_i) {\sf h}^+(k s_i) {\sf Y}(\vns_i) =
{\sf Y}(\vnr_j)^T {\sf h}^+(k r_j) {\sf j}(k s_j) {\sf Y}(\vns_j) \; .
\label{eq:neargreen0}
\end{equation}
One sees here already, since $\vcr$ and $\vcs$ are independent and
arbitrary, the likely proportionality between
${\sf h}^+(k s_i) {\sf Y}(\vns_i)$ and ${\sf j}(k s_j) {\sf Y}(\vns_j)$.
To see this explicitly we first multiply on the left by ${\sf Y}(\vnr_i)$,
integrate over the angular variables $\vnr_i$, and use orthonormality of
spherical harmonics to get
\begin{equation}
{\sf j}(k r_i) {\sf h}^+(k s_i) {\sf Y}(\vns_i) =
\left[ \int {\sf Y}(\vnr_i) {\sf Y}(\vnr_j)^T {\sf h}^+(k r_j)
\; d\Omega_{\vcr_i} \right]
{\sf j}(k s_j) {\sf Y}(\vns_j) \; .
\end{equation}
To eliminate the ${\sf j}(k r_i)$ term, and thus the dependence on
$r_i$ on the left-hand side (lhs), we use the famous Wronskian
formula (\ref{eq:bessneuwr}) for spherical Bessel functions in the form
\begin{equation}
{\sf h}^+(k r) \frac{\partial}{\partial r} \left[ {\sf j}(k r) \right]
- \frac{\partial}{\partial r} \left[ {\sf h}^+(k r) \right] {\sf j}(k r) 
= - \frac{i}{k r^2} {\sf I} \; .
\end{equation}
Thus, if we multiply our main equation on the left by the derivative
with respect to (wrt) $r_i$
of ${\sf h}^+(k r_i)$, and the same derivative of our main equation by
${\sf h}^+(k r_i)$, and take the difference, we obtain, after some
rearrangement,
\begin{eqnarray}
{\sf h}^+(k s_i) {\sf Y}(\vns_i) = i k {\sf K}_f^{j i \; T}  \;
{\sf j}(k s_j) {\sf Y}(\vns_j)
& \mbox{for} & s_j < R_{ij} \; ;
\label{eq:reexpq0neargen}
\end{eqnarray}
where we have defined the important matrix
\begin{equation}
{\sf K}_f^{j i \; T} \equiv r_i^2
\int \left\{ {\sf h}^+(k r_i) {\sf Y}(\vnr_i)
\frac{\partial}{\partial r_i}
\left[ {\sf Y}(\vnr_j)^T {\sf h}^+(k r_j) \right]
- \frac{\partial}{\partial r_i}
\left[ {\sf h}^+(k r_i) \right] {\sf Y}(\vnr_i)
{\sf Y}(\vnr_j)^T {\sf h}^+(k r_j)
\right\} \; d\Omega_{\vcr_i} \; .
\label{eq:kij0defn}
\end{equation}
If we introduce some notation from Ref.~\cite{Na+al.90a} for vectors
of free radial solutions, denoted ${\sf J}(k \vcr)$, ${\sf N}(k \vcr)$,
and ${\sf H}^{\pm}(k \vcr)$, with elements given by
\begin{equation}
J_L(k \vcr) \equiv j_l(k r) Y_L(\vnr) \; ,
\end{equation}
\begin{equation}
N_L(k \vcr) \equiv n_l(k r) Y_L(\vnr) \; ,
\end{equation}
and
\begin{equation}
H_L^{\pm}(k \vcr) \equiv h_l^{\pm}(k r) Y_L(\vnr) \; ,
\end{equation}
then we may write
\begin{eqnarray}
{\sf K}_f^{j i} & = & r_i^2
\int \left\{
\frac{\partial}{\partial r_i}
\left[ {\sf H}^+(k \vcr_j) \right]
{\sf H}^+(k \vcr_i)^T
- {\sf H}^+(k \vcr_j)
\frac{\partial}{\partial r_i}
\left[ {\sf H}^+(k \vcr_i)^T \right]
\right\} \; d\Omega_{\vcr_i} \nonumber \\
& = &
\int_{\partial \tau_i(r_i)} \left\{
\nabla_{\vcr_i}
\left[ {\sf H}^+(k \vcr_j) \right]
{\sf H}^+(k \vcr_i)^T
- {\sf H}^+(k \vcr_j)
\nabla_{\vcr_i}
\left[ {\sf H}^+(k \vcr_i)^T \right]
\right\} \cdot \vnn \; d\sigma
\; ;
\label{eq:kij0alt}
\end{eqnarray}
where the volume $\tau_i(r_i)$ is a sphere of radius $r_i$ centered
on $\vcR_i$.

Now Eq.~(\ref{eq:reexpq0neargen}) is clearly another form of the
near-region re-expansion formula (\ref{eq:reexpq0near}) for the free
irregular solutions, and comparison with this latter shows,
using also Eq.~(\ref{eq:gauntlincom}), that
\begin{equation}
{\sf K}_f^{j i} = -(i/k) {\sf F}(k;\vcR_{ij})
= - (i/k) \xi \; {\sf \Lambda}[4 \pi \xi {\sf H}^+(k \vcR_{i j})] \;
\xi^{\dagger} \; .
\label{eq:kij0closed}
\end{equation}
It should also be clear that ${\sf K}_f^{j i}$ is constant, even though
the definition (\ref{eq:kij0defn}) and alternative forms
(\ref{eq:kij0alt}) appear to depend on $r_i$. We may show this directly
by using Green's theorem to transform the second form in
Eq.~(\ref{eq:kij0alt}) into a volume integral. First, 
however, we note that the gradient operator in the integrand of this
latter form may be taken to be wrt $\vcr$, since $\vcr_i$ is just a
translation of this by a constant vector.
One then sees that this integrand is anti-symmetric
under the simultaneous exchange of the indices $i$ and $j$ and matrix
transposition. Second, from Fig.~\ref{fig-2}
we see that the constraints on our geometric
configuration imply that $r_i < R_{i j} = R_{j i}$ and we can
therefore use our preceding results
(\ref{eq:reexpq0neargen}) to write
\begin{eqnarray}
{\sf h}^+(k r_j) {\sf Y}(\vnr_j) = i k {\sf K}_f^{i j \; T}  \;
{\sf j}(k r_i) {\sf Y}(\vnr_i) \; .
\end{eqnarray}
We may now use this and Eq.~(\ref{eq:reexpq0neargen}) to eliminate the
irregular solutions altogether from Eq.~(\ref{eq:neargreen0}) and find,
on rearranging, that
\begin{equation}
{\sf Y}(\vnr_i)^T {\sf j}(k r_i)
\left[ i k {\sf K}_f^{i j} - i k {\sf K}_f^{j i\; T} \right]
{\sf j}(k s_j) {\sf Y}(\vns_j)
= 0 \; ,
\nonumber
\end{equation}
from which we deduce that
\begin{equation}
{\sf K}_f^{i j} = {\sf K}_f^{j i \; T} \; .
\label{eq:kij0sym}
\end{equation}

Now the integrand of Eq.~(\ref{eq:kij0alt}) is singular at
$\vcr=\vcR_i$ and $\vcr=\vcR_j$. Using Green's theorem
(\ref{eq:greenthm}) we find that
\begin{equation}
{\sf K}_f^{j i} =
\int_{\tau_i(r_i)} \left\{
\left[ (\nabla^2 + E )
{\sf H}^+(k \vcr_j) \right]
{\sf H}^+(k \vcr_i)^T
- {\sf H}^+(k \vcr_j)
\left[ (\nabla^2 + E )
{\sf H}^+(k \vcr_i)^T \right]
\right\} \; d^3r
\; .
\nonumber
\end{equation}
Since the vectors ${\sf H}^+(k \vcr_i)$ and ${\sf H}^+(k \vcr_j)$
are solutions of the free Schr\"{o}dinger equation, the integrand is
identically zero away from the singularities. Thus the only finite
contribution to the integral comes from the singularities
themselves. It is then clear that the surface integral expression
for ${\sf K}_f^{j i}$ may be taken over any surface which
encloses only the singularity at $\vcr=\vcR_i$. In particular,
the integral is independent of $r_i$. The possibility then arises
that the closed form of Eq.~(\ref{eq:kij0closed}) may be obtained by
evaluating the integral as $r_i \rightarrow 0$. This is not so
interesting in the free-electron case, but might be applicable
to our general irregular solutions which we consider below.

It is not difficult to show that the contribution to such an integral
from the singularity at $\vcr=\vcR_j$ would be equal and opposite to
that at $\vcr=\vcR_i$. This may be seen by evaluating the second form
in Eq.~(\ref{eq:kij0alt}) over the surface ${\partial \tau_j(r_j)}$,
using the anti-symmetry of the integrand to show that the result is
$- {\sf K}_f^{i j \; T}$, and noting Eq.~(\ref{eq:kij0sym}).
Alternatively we might consider an integral over a surface which
encloses both singularities and show that the result is zero.
One simple possibility is a spherical surface of radius
$R > R_{i j}$ centered on $\vcr = \vcR_i$, so that we can use
the far-region re-expansion formula (\ref{eq:reexpq0far}) to replace
the term ${\sf H}^+(k \vcr_j)$. Thus we would have, say,
\begin{eqnarray}
{\sf X}^{i j} & \equiv &
\int_{\partial \tau_i(R)} \left\{
\nabla
\left[ {\sf H}^+(k \vcr_j) \right]
{\sf H}^+(k \vcr_i)^T
- {\sf H}^+(k \vcr_j)
\nabla
\left[ {\sf H}^+(k \vcr_i)^T \right]
\right\} \cdot \vnn \; d\sigma \nonumber \\
& = &
{\sf D}(k;\vcR_{ij})
\int_{\partial \tau_i(R)} \left\{
\nabla
\left[ {\sf H}^+(k \vcr_i) \right]
{\sf H}^+(k \vcr_i)^T
- {\sf H}^+(k \vcr_i)
\nabla
\left[ {\sf H}^+(k \vcr_i)^T \right]
\right\} \cdot \vnn \; d\sigma \nonumber \\
& = & 0 \; ,
\nonumber
\end{eqnarray}
where the last step is justified by noting that the integrand is
anti-symmetric and, when the integration over angles is done,
the result is diagonal.

With the foregoing development in mind we may proceed to consider
in a similar way the near-region re-expansion of the general
irregular solutions. Noting again the situation in Fig.~\ref{fig-2}
and the order relations, the expressions for the general Green
function relative to the two centers, $\vcR_i$ and $\vcR_j$,
lead to
\begin{equation}
{\sf Y}(\vnr_i)^T {\sf p}_i(r_i)
\left( {\sf M}_{i +}^{-1} \right)^{T}
{\sf q}^{(+)}_i(s_i)^T {\sf Y}(\vns_i)
= {\sf Y}(\vnr_j)^T {\sf q}^{(+)}_j(r_j)
\left( {\sf M}_{j +}^{-1} \right)
{\sf p}_j(s_j)^T {\sf Y}(\vns_j)  \; .
\end{equation}
As before we have then
\begin{equation}
{\sf p}_i(r_i)
\left( {\sf M}_{i +}^{-1} \right)^{T}
{\sf q}^{(+)}_i(s_i)^T {\sf Y}(\vns_i) =
\left[ \int {\sf Y}(\vnr_i) {\sf Y}(\vnr_j)^T
{\sf q}^{(+)}_j(r_j)
\; d\Omega_{\vcr_i} \right]
\left( {\sf M}_{j +}^{-1} \right)
{\sf p}_j(s_j)^T {\sf Y}(\vns_j) \; .
\nonumber
\end{equation}
Using now Eq.~(\ref{eq:pqwronski}) we obtain
\begin{eqnarray}
{\sf q}^{(+)}_i(s_i)^T {\sf Y}(\vns_i) = {\sf K}^{j i \; T}  \;
\left( {\sf M}_{j +}^{-1} \right)
{\sf p}_j(s_j)^T {\sf Y}(\vns_j)
& \mbox{for} & s_j < R_{ij} \; ;
\label{eq:reexpqneargen}
\end{eqnarray}
with the definition
\begin{eqnarray}
\lefteqn{{\sf K}^{j i \; T} \equiv } \nonumber \\
 & & - r_i^2
\int \left\{ {\sf q}^{(+)}_i(r_i)^T {\sf Y}(\vnr_i)
{\sf Y}(\vnr_j)^T 
\frac{\partial}{\partial r_i}
\left[ {\sf q}^{(+)}_j(r_j) \right]
- \frac{\partial}{\partial r_i}
\left[ {\sf q}^{(+)}_i(r_i)^T \right] {\sf Y}(\vnr_i)
{\sf Y}(\vnr_j)^T {\sf q}^{(+)}_j(r_j)
\right\} \, d\Omega_{\vcr_i} . \nonumber \\
\label{eq:kijdefn}
\end{eqnarray}
One may easily check the correspondence with our earlier formula in the
limit of zero potential.
Again we may consider the re-expansion from center $i$ to center $j$ and 
obtain
\begin{eqnarray}
{\sf q}^{(+)}_j(r_j)^T {\sf Y}(\vnr_j) = {\sf K}^{i j \; T}  \;
\left( {\sf M}_{i +}^{-1} \right)
{\sf p}_i(r_i)^T {\sf Y}(\vnr_i)
& \mbox{for} & r_i < R_{ij} \; ;
\end{eqnarray}
using the two re-expansion formulas to prove the symmetry
\begin{equation}
{\sf K}^{i j} = {\sf K}^{j i \; T} \; .
\label{eq:kijsym}
\end{equation}
Let us define vectors of general solutions by
\begin{equation}
{\sf P}(\vcr) \equiv {\sf p}(r)^T {\sf Y}(\vnr) \; ,
\end{equation}
and
\begin{equation}
{\sf Q}^{(\pm)}(\vcr) \equiv {\sf q}^{(\pm)}(r)^T {\sf Y}(\vnr) \; .
\end{equation}
We may write
\begin{equation}
{\sf K}^{j i} = - {r_i}^2
\int_{\partial \tau_i(r_i)} \left\{
\nabla
\left[ {\sf Q}_j^{(+)}(\vcr_j) \right]
{\sf Q}_i^{(+)}(\vcr_i)^T
- {\sf Q}_j^{(+)}(\vcr_j) \;
\nabla
\left[ {\sf Q}_i^{(+)}(\vcr_i)^T \right]
\right\} \cdot \vnn \; d\sigma
\; ,
\label{eq:kijalt}
\end{equation}
and show, as above, the independence wrt $r_i$. Precisely the
same comments as before, about surface and volume integrals,
apply in this case, except that, in using Green's theorem
it should be noted that the terms like
$(\nabla^2 + E){\sf Q}_j^{(+)}(\vcr_j)$ are not zero, but
$V_j(\vcr_j){\sf Q}_j^{(+)}(\vcr_j)$, and they will cancel
each other out in the integrand of the volume integral
(since $V_i(\vcr_i) = V_j(\vcr_j) = V(\vcr)$ by definition),
so that this is again zero away from the singularities.

In the absence of a closed formula for the ${\sf K}^{j i}$ we must evaluate
them numerically via Eq.~(\ref{eq:kijalt}). The freedom in choice of surface
may be invoked to use the atomic-sphere surfaces of radius $b_i$ that we
introduced earlier. The possibility that a closed form might be obtained by
taking the limit as $r_i \rightarrow 0$ remains to be investigated.


\subsection{Multiple-scattered-wave method}

Having developed the necessary new mathematical machinery we may now return
to consideration of the multiple-scattering problem. Here we follow the
general approach of Ref.~\cite{Na+al.86a}, but with several important
differences. It will be clear that the foregoing development of
re-expansion formulas for general solutions of Schr\"{o}dinger's equation
is intended for the distorting potential $V_I(\vcr)$. Thus it is assumed
that we have available, for each atomic center with position vector
$\vcR_i$ (for $i=1,\cdots,N$), the exact solution matrices
${\sf p}_i$ and ${\sf q}^{(+)}_i$, regular at $\vcr_i=0$ and $\infty$,
respectively, in this potential. Finding these solutions is thus, 
of course, a large part of the computational burden in our approach
to the multiple-scattering problem. For each center we evaluate the
constant Wronskian matrix ${\sf M}_{i +}$ from these solutions, via
Eq.~(\ref{eq:pqwronski}), at some suitable radius $r_i=b_i$, say.

We are interested in finding the scattering solution $\psi^+(\vcr)$ of the
Schr\"{o}dinger equation for our multicenter potential $V(\vcr)$, which
develops from an incoming plane wave $\phi_0(\vcr)$. It is assumed that we
have solved the scattering problem for our potential $V_I(\vcr)$ in terms
of the ${\sf p}_i$, ${\sf q}^{(+)}_i$, and ${\sf M}_{i +}$ just mentioned,
and thus have expressions for the Green function $G_I(\vcr,\vcs)$, of the
form of Eq.~(\ref{eq:gfinal}), with respect to any atomic center $i$.
In particular, we have an explicit expression for the distorted wave
$\chi^+(\vcr)$ given by Eq.~(\ref{eq:chip})

Following our earlier discussion it is then clear that to find
$\psi^+(\vcr)$ we must solve the Lippmann-Schwinger equation
(\ref{eq:lippschwdw}) with the Green function $G_I(\vcr,\vcs)$ and the
singular part $V_A(\vcr)$ of our multicenter potential, together with
the distorted wave as inhomogeneous part, i.e.,
\begin{equation}
\psi^+(\vcr) = \chi^+(\vcr) + \int G_I^+(\vcr,\vcs) V_A(\vcs)
\psi^+(\vcs) \; d^3\!s \, .
\label{eq:lipschmsw}
\end{equation}
As usual the first step is to split the volume integral in this equation
according to the molecular partition. Therefore we write
\begin{equation}
\psi^+(\vcr) = \chi^+(\vcr) + \sum_{i=1}^N
\int_{\tau_i} G_I^+(\vcr,\vcs) V_A(\vcs) \psi^+(\vcs) \; d^3\!s
+ \int_{\cal I} G_I^+(\vcr,\vcs) V_A(\vcs) \psi^+(\vcs) \; d^3\!s \, ,
\nonumber
\end{equation}
where we have denoted the interstitial volume by ${\cal I}$, so that
\begin{equation}
{\cal I} \equiv \overline{\bigcup_{i=1}^N \tau_i} \; .
\end{equation}
Now, by construction, the singular part of the potential is zero in the
interstitial region, so the last term in our equation vanishes,
and we therefore have
\begin{equation}
\psi^+(\vcr) = \chi^+(\vcr) + \sum_{i=1}^N
\int_{\tau_i} G_I^+(\vcr,\vcs) V_A(\vcs) \psi^+(\vcs) \; d^3\!s \, .
\nonumber
\end{equation}
Now $\psi^+(\vcr)$ is a solution of the Schr\"{o}dinger equation
(\ref{eq:schreq}) for the full potential, and because of
Eq.~(\ref{eq:potsplit}) we have
\begin{equation}
V_A(\vcr) \psi^+(\vcr) =
\left[ {\nabla}^2 + E - V_I(\vcr) \right] \psi^+(\vcr) \; .
\end{equation}
Using this we can eliminate $V_A(\vcr)$ and obtain
\begin{eqnarray}
\psi^+(\vcr) & = &
\chi^+(\vcr) + \sum_{i=1}^N
\int_{\tau_i} G_I^+(\vcr,\vcs)
\left[ \left({\nabla}_{\vcs}^2 + E \right) \psi^+(\vcs) - 
V_I(\vcs) \psi^+(\vcs) \right] \; d^3\!s \nonumber \\
& = &
\chi^+(\vcr) + \sum_{i=1}^N
\int_{\tau_i}
\left\{
\left[ G_I^+(\vcr,\vcs)
\left({\nabla}_{\vcs}^2 + E \right) \psi^+(\vcs)
\right. \right. \nonumber \\
& &
\left. \left.
- \psi^+(\vcs) \left({\nabla}_{\vcs}^2 + E \right) 
G_I^+(\vcr,\vcs) \right]
+ \delta^3(\vcr - \vcs) \psi^+(\vcs) \right\} \; d^3\!s \, ,
\end{eqnarray}
where in the last step we have used the basic property (\ref{eq:gdef})
of the Green function, noting the symmetry (\ref{eq:gvarsym}). Because
of the delta-function term we see that there are two distinct cases
to be considered. Invoking first Green's theorem, we find that
\begin{eqnarray}
\psi^+(\vcr) =
\chi^+(\vcr) + \sum_{i=1}^N
\int_{\partial \tau_i}
\left[ G_I^+(\vcr,\vcs)
{\nabla}_{\vcs} \psi^+(\vcs) - 
\psi^+(\vcs) {\nabla}_{\vcs} 
G_I^+(\vcr,\vcs) \right] \cdot \vnn_i \; d\sigma_i \;
& & \mbox{if} \; \vcr \notin \bigcup_{i=1}^N \tau_i
\, , \nonumber \\
\label{eq:lipschsurfinterst}
\end{eqnarray}
and
\begin{eqnarray}
0 =
\chi^+(\vcr) + \sum_{i=1}^N
\int_{\partial \tau_i}
\left[ G_I^+(\vcr,\vcs)
{\nabla}_{\vcs} \psi^+(\vcs) - 
\psi^+(\vcs) {\nabla}_{\vcs} 
G_I^+(\vcr,\vcs) \right] \cdot \vnn_i \; d\sigma_i \;
& & \mbox{if} \; \vcr \in \bigcup_{i=1}^N \tau_i
\, . \nonumber \\
\label{eq:lipschsurfatom}
\end{eqnarray}
We may now rewrite these equations in terms of radial
Schr\"{o}dinger-equation solutions at each atomic center.
We have already discussed the
Green function at considerable length, and the associated radial
functions may be taken as known. We require also expressions for the
wavefunction $\psi^+(\vcr)$ inside each atomic sphere $\tau_i$. Here we
follow the development of Natoli, Benfatto, and Doniach \cite{Na+al.86a}
who show that we may write, for $\vcr \in \tau_i$,
\begin{equation}
\psi^+(\vcr) = \sum_{L L'} C_{L'}^i R_{L L'}^i(r_i) Y_L(\vnr_i) \; ,
\label{eq:wvfnatom}
\end{equation}
where the $C_{L'}^i$ are constants, and the $R_{L L'}^i(r_i)$
constitute a set, indexed by $L'$, of linearly independent solutions
to the radial Schr\"{o}dinger equations at center $i$,
\begin{equation}
\left[ \frac{1}{r_i^2} \left( \frac{\partial}{\partial r_i}
\left( r_i^2 \frac{\partial}{\partial r_i} \right) \right)
- \frac{l(l+1)}{r_i^2} + E
\right] R_{L L'}^i(r_i) 
 - \sum_{L''} v_{L L''}^i(r_i) R_{L'' L'}^i(r_i)  = 0 \; ,
\label{eq:coupradschreqatom}
\end{equation}
which are regular at the origin. The potential matrix is given by
\begin{equation}
v^i_{L L'}(r_i) \equiv \sum_{L''} I(L,L',L'') V^i_{L''}(r_i) \; ,
\label{eq:potmxatom}
\end{equation}
where $V^i_{L}(r_i)$ comes from the partial-wave expansion of the
{\it full} molecular potential $V(\vcr)$ at center $i$, i.e.,
\begin{equation}
V(\vcr) \equiv \sum_L V^i_L(r_i) Y_L(\vnr_i) \; .
\label{eq:vshxatom}
\end{equation}
In our matrix notation we may write the atomic solution as
\begin{equation}
\psi^+(\vcr) = {\sf Y}(\vnr_i)^T {\sf R}^i(r_i) {\sf C}^i \; .
\end{equation}

If we now consider the case where our general vector $\vcr$ is
inside one of the atomic spheres and return to the second of our
main equations, Eq.~(\ref{eq:lipschsurfatom}), then we see that
there are two possibilities: either the integration variable is
on the surface of the same sphere, or another one. 
For the first of these possibilities we have $s_i > r_i$ and we
can therefore write the Green function as
\begin{equation}
G_I^+(\vcr,\vcs) = {\sf Y}(\vnr_i)^T
{\sf p}_i(r_i) \left( {\sf M}_{i +}^{-1} \right)^{T}
{\sf q}_i^{(+)}(s_i)^T {\sf Y}(\vns_i) \; .
\nonumber
\end{equation}
The surface integral may be then evaluated as follows:
\begin{eqnarray*}
\lefteqn{\int_{\partial \tau_i}
\left[ G_I^+(\vcr,\vcs)
{\nabla}_{\vcs} \psi^+(\vcs) - 
\psi^+(\vcs) {\nabla}_{\vcs} 
G_I^+(\vcr,\vcs) \right]
\cdot \vnn_i \; d\sigma_i} \\
& = & {\sf Y}(\vnr_i)^T {\sf p}_i(r_i)
\left( {\sf M}_{i +}^{-1} \right)^{T}
\int_{\partial \tau_i}
\left[ 
{\sf q}_i^{(+)}(s_i)^T {\sf Y}(\vns_i)
{\sf Y}(\vns_i)^T {{\sf R}^i}'(s_i) {\sf C}^i 
\right. \\
& &
\left.
- {{\sf q}_i^{(+)}}'(s_i)^T {\sf Y}(\vns_i)
{\sf Y}(\vns_i)^T {{\sf R}^i}(s_i) {\sf C}^i 
\right] \; d\sigma_i \\
& = & {b_i}^2 {\sf Y}(\vnr_i)^T {\sf p}_i(r_i)
\left( {\sf M}_{i +}^{-1} \right)^{T}
\left[ 
{\sf q}_i^{(+)}(b_i)^T {{\sf R}^i}'(b_i)
- {{\sf q}_i^{(+)}}'(b_i)^T {{\sf R}^i}(b_i) 
\right] {\sf C}^i \; .
\end{eqnarray*}
For the second possibility, with $j \neq i$, we have $r_j > s_j$,
and the Green function is given by
\begin{equation}
G_I^+(\vcr,\vcs) = {\sf Y}(\vnr_j)^T
{\sf q}_j^{(+)}(r_j)
\left( {\sf M}_{j +}^{-1} \right)
{\sf p}_j(s_j)^T
{\sf Y}(\vns_j) \; ,
\nonumber
\end{equation}
which becomes, noting that $r_i < R_{ij}$ and using the 
re-expansion formula (\ref{eq:reexpqneargen}), 
\begin{equation}
G_I^+(\vcr,\vcs) =
{\sf Y}(\vnr_i)^T
{\sf p}_i(r_i)
\left( {\sf M}_{i +}^{-1} \right)^T \, {\sf K}^{i j} \,
\left( {\sf M}_{j +}^{-1} \right)
{\sf p}_j(s_j)^T
{\sf Y}(\vns_j) \; .
\end{equation}
We then have
\begin{eqnarray*}
\lefteqn{\int_{\partial \tau_j}
\left[ G_I^+(\vcr,\vcs)
{\nabla}_{\vcs} \psi^+(\vcs) - 
\psi^+(\vcs) {\nabla}_{\vcs} 
G_I^+(\vcr,\vcs) \right]
\cdot \vnn_j \; d\sigma_j} \\
& = & {b_j}^2 {\sf Y}(\vnr_i)^T {\sf p}_i(r_i)
\left( {\sf M}_{i +}^{-1} \right)^{T} \, {\sf K}^{i j} \,
\left( {\sf M}_{j +}^{-1} \right)
\left[ 
{\sf p}_j(b_j)^T {{\sf R}^j}'(b_j)
- {{\sf p}_j}'(b_j)^T {{\sf R}^j}(b_j) 
\right] {\sf C}^j \; .
\end{eqnarray*}

If we define the Wronskian of two matrix functions
${\sf a}(r)$ and ${\sf b}(r)$, say, by
\begin{equation}
{\sf W}[{\sf a},{\sf b}](r) \equiv
{\sf a}(r) {\sf b}'(r) - {\sf a}'(r) {\sf b}(r) \; ,
\end{equation}
then our results may be written
\begin{eqnarray}
\lefteqn{0 =} \nonumber \\
& & \chi^+(\vcr) +
{\sf Y}(\vnr_i)^T
{\sf p}_i(r_i)
\left( {\sf M}_{i +}^{-1} \right)^T \!
\left\{
{b_i}^2 {\sf W}[ {{\sf q}_i^{(+)}}^T \! ,
{{\sf R}^i} ](b_i) {\sf C}^i +
\sum_{j \neq i} 
{b_j}^2\, {\sf K}^{i j} \,
\left( {\sf M}_{j +}^{-1} \right)
{\sf W}[ {{\sf p}_j}^T , {{\sf R}^j} ](b_j)
{\sf C}^j 
\right\} . \nonumber \\
\end{eqnarray}
Let us now define
\begin{equation}
{\sf B}^i \equiv {b_i}^2
\left( {\sf M}_{i +}^{-1} \right)
{\sf W}[ {{\sf p}_i}^T , {{\sf R}^i} ](b_i)
{\sf C}^i \; ,
\end{equation}
which we may presumably invert to recover
\begin{equation}
{\sf C}^i = {b_i}^{-2} \,
{\sf W}[ {{\sf p}_i}^T , {{\sf R}^i} ](b_i)^{-1}
{\sf M}_{i +} {\sf B}^i \; .
\label{eq:wvfnatomcoeff}
\end{equation}
Define also, for each atomic site, the matrices
\begin{equation}
{{\sf T}_a^i}^{-1} \equiv
{\sf W}[ {{\sf q}_i^{(+)}}^T , {{\sf R}^i} ](b_i) \;
{\sf W}[ {{\sf p}_i}^T , {{\sf R}^i} ](b_i)^{-1} \;
{\sf M}_{i +} \; .
\label{eq:tmatatomdef}
\end{equation}
We now note that the inhomogeneous term may be written, using
Eqs. (\ref{eq:chip}) and (\ref{eq:reexpnpij}) (noting that the
re-expansion we wish to effect is from the origin of coordinates
to center $i$), as
\begin{equation}
\chi^+(\vcr) =
(4 \pi / k)\,
{\sf Y}(\vnr_i)^T {\sf p}_i(r_i)
\left( {\sf M}_{i +}^{-1} \right)^{T} 
{\sf D}(k;\vcR_{i}) \,
\xi \, {\sf Y}(\vnk) \; .
\end{equation}
We may now write our results
\begin{equation}
0 =
{\sf Y}(\vnr_i)^T
{\sf p}_i(r_i)
\left( {\sf M}_{i +}^{-1} \right)^T
\left\{
(4 \pi / k)\,
{\sf D}(k;\vcR_{i}) \,
\xi {\sf Y}(\vnk) 
+ {{\sf T}_a^i}^{-1} {\sf B}^i
+ \sum_{j \neq i} 
\, {\sf K}^{i j} \,{\sf B}^j 
\right\} \, ;
\end{equation}
for which it is necessary that
\begin{equation}
{{\sf T}_a^i}^{-1} {\sf B}^i
+ \sum_{j \neq i} 
\, {\sf K}^{i j} \,{\sf B}^j
= - (4 \pi / k)\,
{\sf D}(k;\vcR_{i}) \,
\xi \, {\sf Y}(\vnk)  \, .
\label{eq:msecont}
\end{equation}
There will be one of these equations for each $i$, and together
they constitute the multiple-scattering equations for the case of an
incoming plane wave of direction $\vnk$. Clearly they form a linear
system which may be solved for the unknown vectors ${\sf B}^i$ by
inversion of the secular matrix whose elements are defined by
\begin{equation}
S^{i j}_{L L'} \equiv {T_a^i}^{-1}_{L L'} \delta_{i j}
+ (1 - \delta_{i j}) K^{i j}_{L L'} \; .
\label{eq:secmtx}
\end{equation}
More generally, following Ref.~\cite{LloSmi72a}, we replace the exciting
amplitude $4 \pi i^l Y_L(\vnk)$ by $\delta_{L L''}$ and find a solution
vector $B^i_L(L'')$ for each partial-wave channel $L''$. The results may
be then applied to more general situations.

With the solution vector ${\sf B}^i$ we may find the wavefunction inside
the atomic spheres by using Eq.~(\ref{eq:wvfnatomcoeff}) to
determine the coefficients ${\sf C}^i$, then inserting them into
Eq.~(\ref{eq:wvfnatom}). For the interstitial-region
wavefunction we must return to perform the surface
integrals in Eq.~(\ref{eq:lipschsurfinterst}).
In this case, for $\vcr$ in the
interstitial region and $\vcs$ on the surface of atomic sphere $i$,
we have $r_i > s_i$, so that
\begin{equation}
G_I^+(\vcr,\vcs) = {\sf Y}(\vnr_i)^T
{\sf q}_i^{(+)}(r_i)
\left( {\sf M}_{i +}^{-1} \right)
{\sf p}_i(s_i)^T
{\sf Y}(\vns_i) \; .
\nonumber
\end{equation}
We obtain then
\begin{eqnarray}
\psi^+(\vcr) & = &
\chi^+(\vcr) + \sum_{i=1}^N
{b_i}^2  {\sf Y}(\vnr_i)^T
{\sf q}_i^{(+)}(r_i)
\left( {\sf M}_{i +}^{-1} \right)
\left[ 
{\sf p}_i(b_i)^T {{\sf R}^i}'(b_i)
- {{\sf p}_i}'(b_i)^T {{\sf R}^i}(b_i) 
\right] {\sf C}^i \nonumber \\
& = &
\chi^+(\vcr) + \sum_{i=1}^N
{\sf Y}(\vnr_i)^T
{\sf q}_i^{(+)}(r_i) {\sf B}^i \; .
\end{eqnarray}
This completes the solution of the multiple-scattering problem.


\subsubsection{Symmetry of the secular matrix}

We see from Eq.~(\ref{eq:kijsym}) that the second term in the secular
matrix (\ref{eq:secmtx}) is symmetric. The question then arises as to
whether the first term ${{\sf T}_a^i}^{-1}$ has any symmetry. There
appears to be none manifest in its definition (\ref{eq:tmatatomdef}).
However, the author has treated (in Ref.~\cite{Foulis88a}) a similar
problem for the bound-state version of the method of Natoli, Benfatto,
and Doniach \cite{Na+al.86a}. In fact, a proof of the symmetry of
${{\sf T}_a^i}^{-1}$ may also be developed in our present case.

To see this we first define the matrices ${\sf d}$ and ${\sf t}$ by
\begin{equation}
{\sf d} \equiv {\sf W}[ {{\sf p}}^T , {{\sf R}} ](b)^{-1} \;
{\sf M}_{+} \; ,
\nonumber
\end{equation}
and
\begin{equation}
{\sf t} \equiv ({\sf d}^{-1})^T \,{{\sf T}_a}^{-1}\,
({\sf d}^{-1}) \; ,
\nonumber
\end{equation}
suppressing for the moment the atomic index $i$.
We may recover our original matrix via
\begin{equation}
{{\sf T}_a}^{-1} = {\sf d}^T \,{\sf t} \, {\sf d} \; .
\nonumber
\end{equation}
It is clear from the definitions that the symmetry of ${\sf t}$
is equivalent to that of ${{\sf T}_a}^{-1}$.

Expanding ${\sf t}$ we have (suppressing dependence on the radius $b$)
\begin{eqnarray}
{\sf t} & = &
{\sf W}[ {{\sf p}}^T , {{\sf R}} ]^{T} \,
({\sf M}_{+}^{-1})^T \,
{\sf W}[ {{\sf q}^{(+)}}^T , {{\sf R}} ] \nonumber \\
& = &
\left\{ {{\sf R}'}^T \left[ {\sf p}
\,({\sf M}_{+}^{-1})^T\,
{{\sf q}^{(+)}}^T \right] {\sf R}' \right\}
+
\left\{ {{\sf R}}^T \left[ {\sf p}'
\,({\sf M}_{+}^{-1})^T\,
{{{\sf q}^{(+)}}'}^T \right] {\sf R} \right\}
- {\sf f} \; ,
\label{eq:teqsymsymf}
\end{eqnarray}
where we have defined
\begin{equation}
{\sf f} \equiv
{{\sf R}'}^T \left[ {\sf p}
\,({\sf M}_{+}^{-1})^T\,
{{{\sf q}^{(+)}}'}^T \right] {\sf R}
+ {{\sf R}}^T \left[ {\sf p}'
\,({\sf M}_{+}^{-1})^T\,
{{\sf q}^{(+)}}^T \right] {\sf R}'
\label{eq:fequivstuff}
\end{equation}
It may be seen now that the two expressions collected
between braces in Eq.~(\ref{eq:teqsymsymf})
are each individually symmetric; the first by virtue of
Eq.~(\ref{eq:gpartcont}) and the second from
Eq.~(\ref{eq:gpartd1d1}). So the symmetry of ${\sf f}$
is equivalent to that of ${\sf t}$.
Substituting from Eq.~(\ref{eq:gpartinhom}) in the second term of
Eq.~(\ref{eq:fequivstuff}) leads to
\begin{equation*}
{\sf f} =
\left\{ {{\sf R}'}^T \left[ {\sf p}
\,({\sf M}_{+}^{-1})^T\,
{{{\sf q}^{(+)}}'}^T \right] {\sf R}
+ {{\sf R}}^T \left[ {{\sf q}^{(+)}}'
\,({\sf M}_{+}^{-1})\,
{\sf p}^T \right] {\sf R}' \right\} 
- \frac{1}{b^2} {{\sf R}}^T {\sf R}' \; .
\end{equation*}
The term in braces here is now manifestly symmetric and we are left
with the term in ${{\sf R}}^T {\sf R}'$.
However this last is also symmetric for similar reasons to
the comparable term in ${\sf p}$ of Eq.~(\ref{eq:ppsym}), i.e.,
since ${\sf R}$ is the matrix solution, regular at the origin,
of the coupled radial Schr\"{o}dinger equations for a real potential.
The proof may be seen in Ref.~\cite{Foulis04a}.
Thus our result is shown.


\subsubsection{Bound States}

As mentioned above our new MSW method is readily adapted to bound states.
In their exposition of a non-MT MSW method, Natoli, Benfatto, and
Doniach~\cite{Na+al.86a} do not treat this case explicitly (this is done
in Ref.~\cite{Foulis88a}), but observe that it is achieved by dropping
the inhomogeneous term in the Lippmann-Schwinger equation; making the
analytic continuation $k \rightarrow ik$, $E \rightarrow -E$ where
necessary; using the appropriate re-expansion formulas for modified
spherical Bessel, Neumann, and Hankel functions; and imposing
decaying-wave boundary conditions in the asymptotic region. In our case
we expect that similar comments should apply.

Although again we shall not develop the bound-state case explicitly,
it is necessary to address briefly several issues which arise in our
present approach. Firstly, we note that dropping the inhomogeneous term
in the Lippmann-Schwinger equation (\ref{eq:lipschmsw}) leads to
\begin{equation}
\psi(\vcr) = \int G_I^+(\vcr,\vcs) V_A(\vcs)
\psi(\vcs) \; d^3\!s \,
\label{eq:lipschmswbs}
\end{equation}
as the main equation to be satisfied by a bound state
$\psi(\vcr)$ at some energy $E < 0$. It is not difficult to show
directly that such a $\psi(\vcr)$ will satisfy Schr\"{o}dinger's
equation for the full potential $V(\vcr)$, even though the
two-stage picture of the distorted-wave formalism is lost.

A more substantial problem arises when we come to derive
the re-expansion formulas. In particular,
since there is no inhomogeneous term in Eq.~(\ref{eq:lipschmswbs}),
the argument leading to the re-expansion formula (\ref{eq:reexpnpij})
for the general regular solutions is no longer available to us.
In principle, following the comments above, we may analytically
continue Eq.~(\ref{eq:reexpnpij}) to obtain a similar equation:
\begin{equation}
{\sf Y}(\vnr_i)^T {\sf p}_i(r_i)
\left( {\sf M}_{i}^{-1} \right)^{T} =
{\sf Y}(\vnr_j)^T {\sf p}_j(r_j)
\left( {\sf M}_{j}^{-1} \right)^{T}
{\sf D}^{(-)}(\kappa;\vcR_{ij}) \, ;
\label{eq:reexpnpijeneg}
\end{equation}
where $\kappa$ is such that $\kappa > 0$ and $\kappa^2 = -E$,
and ${\sf D}^{(-)}(\kappa;\vcR_{ij})$ is obtained from 
${\sf D}(k;\vcR_{ij})$ by the substitution $k \rightarrow i\kappa$,
therefore having components
\begin{equation}
D^{(-)}_{L L'}(\kappa;\vcR) = 4 \pi \sum_{L''} i^{l'-l}
i_{l''}(\kappa R) Y_{L''}(\vnR) I(L,L',L'') \; .
\label{eq:dmxenegel}
\end{equation}
The ${\sf p}_i(r_i)$ and ${\sf q}_i(r_i)$, etc., are solution
matrices at our negative energy $E$, and 
the ${\sf q}_i(r_i)$ are asymptotic to the decaying-wave
free solutions ${\sf q}_f(r_i)$, given by
\begin{equation}
{\sf q}_f(r) = i \xi \, {\sf k}^+(\kappa r) \; ,
\label{eq:qfreenegen}
\end{equation}
rather than those of Eq.~(\ref{eq:qnegfree}), since this is the
analytic continuation of the free positive-energy solutions
(\ref{eq:qplusfree}). Such a procedure, however, should really be
more rigorously mathematically justified.

An alternative derivation may be developed via an argument more in
line with our earlier approach. We first note that, taking
${\sf p}_f$ from Eq.~(\ref{eq:pnegfree}) and ${\sf q}_f$ from
Eq.~(\ref{eq:qfreenegen}), we find the inverse transpose of the
constant matrix ${\sf M}_f$ to be
\begin{equation}
\left( {\sf M}_{f}^{-1} \right)^{T}
= i \kappa \, \xi \; .
\end{equation}
The re-expansion formulas for the free solutions at negative energies
are well known (see Eqs. II.25 and II.23 of Ref.~\cite{Johnso73a} for
example). They may be also obtained by the substitution
$k \rightarrow i\kappa$ in our earlier Eqs. (\ref{eq:reexpnp0}) and
(\ref{eq:reexpq0far}). It is not difficult to check that they can be
written in our present notation as
\begin{equation}
{\sf Y}(\vnr_i)^T {\sf p}_f(r_i)
\left( {\sf M}_{f}^{-1} \right)^{T} =
{\sf Y}(\vnr_j)^T {\sf p}_f(r_j)
\left( {\sf M}_{f}^{-1} \right)^{T}
{\sf D}^{(-)}(\kappa;\vcR_{ij})
\label{eq:reexpnp0eneg}
\end{equation}
and 
\begin{eqnarray}
{\sf Y}(\vnr_i)^T {\sf q}_f(r_i)
= {\sf Y}(\vnr_j)^T {\sf q}_f(r_j)
{\sf D}^{(-)}(\kappa;\vcR_{ij})
& \mbox{for} & r_j > R_{ij} \; .
\label{eq:reexpq0fareneg}
\end{eqnarray}
If we now consider a situation such as that of Fig.~\ref{fig-1},
with no restriction on $r_i$ other than that we maintain the order
relations $s_j > R_{ij}$, $s_i > r_i$, and $s_j > r_j$, then we
may use the general Green-function formula to write,
similarly to Eq.~(\ref{eq:gfnqfar}),
\begin{equation}
{\sf Y}(\vnr_i)^T {\sf p}_i(r_i)
\left( {\sf M}_{i}^{-1} \right)^{T}
{\sf q}_i(s_i)^T {\sf Y}(\vns_i)
= {\sf Y}(\vnr_j)^T {\sf p}_j(r_j)
\left( {\sf M}_{j}^{-1} \right)^{T}
{\sf q}_j(s_j)^T {\sf Y}(\vns_j)  \; .
\label{eq:gfnfig1eneg}
\end{equation}
If, for the moment, we assume that the potential is identically zero
beyond some large radius $R$, and if we take the sphere of possible
vectors $\vcs$, large enough to be entirely outside this radius,
then ${\sf q}_j(s_j)$ and ${\sf q}_i(s_i)$ may be taken
to be identically ${\sf q}_f(s_j)$ and ${\sf q}_f(s_i)$ 
respectively in this region. We may therefore use the transpose 
of Eq.~(\ref{eq:reexpq0fareneg}) to substitute on the lhs
of Eq.~(\ref{eq:gfnfig1eneg}), rearranging then to obtain
\begin{equation}
\left[ {\sf Y}(\vnr_i)^T {\sf p}_i(r_i)
\left( {\sf M}_{i}^{-1} \right)^{T}
{\sf D}^{(-)}(\kappa;\vcR_{ij})^{T}
- {\sf Y}(\vnr_j)^T {\sf p}_j(r_j)
\left( {\sf M}_{j}^{-1} \right)^{T} \right]
{\sf q}_f(s_j)^T {\sf Y}(\vns_j) = 0 \; ,
\nonumber
\end{equation}
and we may conclude that the expression in square brackets here
is identically zero. Now by  an argument similar
to that leading to Eq.~(\ref{eq:dkrinv})
we may show that 
\begin{equation}
{\sf D}^{(-)}(\kappa;\vcR)^{-1} = {\sf D}^{(-)}(\kappa;-\vcR) \; .
\label{eq:dkrneginv}
\end{equation}
Furthermore it is easily seen from the explicit formula
(\ref{eq:dmxenegel}) that
\begin{equation}
{\sf D}^{(-)}(\kappa;-\vcR) = {\sf D}^{(-)}(\kappa;\vcR)^*
= {\sf D}^{(-)}(\kappa;\vcR)^{T}
\label{eq:dkrnegprops}
\end{equation}
(showing incidentally that ${\sf D}^{(-)}(\kappa;\vcR)$
is hermitian), and we may therefore deduce that the re-expansion
formula (\ref{eq:reexpnpijeneg}) for the general regular solutions
at negative energies is indeed correct. Presumably we may use
this result to obtain, in a similar way to the derivation of
Eq.~(\ref{eq:reexpqfar}), the corresponding general far-region 
irregular-solution re-expansion formula
\begin{eqnarray}
{\sf Y}(\vns_i)^T {\sf q}_i(s_i) =
{\sf Y}(\vns_j)^T {\sf q}_j(s_j) {\sf D}^{(-)}(\kappa;\vcR_{ij})
& \mbox{for} & s_j > R_{ij} \; .
\label{eq:reexpnqijeneg}
\end{eqnarray}
Our derivation here is less general than we should like because of the
assumption that the potential is zero beyond some large radius.
If, however, we note the fact that
the regular solutions ${\sf p}_i(r_i)$ are
determined by the choice of boundary conditions at $r_i = 0$, and that
the ${\sf D}^{(-)}(\kappa;\vcR)$ are completely given quantities, then
the effect of the potential cutoff will only be felt by the 
${\sf q}_i(r_i)$ and consequently the ${\sf M}_{i}$. Thus, since
we may keep $r_i$ and $r_j$ fixed in Eq.~(\ref{eq:reexpnpijeneg}),
only ${\sf M}_{i}$ and ${\sf M}_{j}$
will vary as a function of the potential
cutoff, and we may plausibly invoke a limiting argument to justify
the formula as the cutoff tends to infinitely large distance,
i.e., no cutoff.

A further issue arises when we consider the potential $V_I(\vcr)$.
It is possible (indeed likely for large molecules) that $V_I(\vcr)$
itself may support bound states. This would appear to cause problems
when our energy passes through that of such a state.
In the bound-state multiple-scattering case the secular 
equation is constructed in a similar way to that for the continuum 
case (\ref{eq:msecont}), except that there is no inhomogeneous
part. The condition for a bound state is that the secular matrix
becomes singular, allowing a non-zero solution vector from which the
corresponding wavefunction may be constructed. As shown in
Ref.~\cite{Foulis04a} a bound state of $V_I(\vcr)$ leads to the
constant matrices ${\sf M}_{i}$, involved in the construction
of the secular matrix, becoming themselves singular,
and would give rise to problems when the
inverses of these matrices are required.
Conceivably it may be possible to factor out such a singularity,
or deal with the problem in another way. This issue
remains to be investigated in more detail.

A systematic way of bypassing this problem is suggested by a comment
of Natoli, Benfatto, and Doniach in the context of their non-MT MSW
method \cite{Na+al.86a}. In particular, we combine their subtraction
of a constant value from the interstitial-region potential
(useful in their case to obtain a convergent
Born expansion for the interstitial-region scattering $T$ matrix)
with the use of an outer sphere (as we describe in the next section)
to obtain a $V_I(\vcr)$ which does not bind.

A final issue is that of normalization of the wavefunction. Since the
wavefunction is constructed from the coefficients ${\sf B}^i$ in the
solution vector of the homogeneous bound-state secular equation, 
the normalization is not known. One must integrate the square of the
resulting function over all space to determine the multiplying
coefficient which leads to unit norm.
In practice one may evaluate this numerically
(although there are analytic methods that may be adapted to our present
approach \cite{Na+al.80a}) and we shall leave further consideration of
this issue outside the scope of the present work.


\subsubsection{Outer Sphere}

As in previous versions of the MSW method it is sometimes useful to add
to the partition of space the outer sphere $\tau_0$,
containing all the atomic spheres,
that we have defined in an earlier section. 
In the region $\overline{\tau_0}$ outside the outer sphere the potential
is expanded as a multipole expansion relative to the sphere center 
$\vcR_0$ and the coupled radial Schr\"{o}dinger equations are solved,
similarly to the atomic centers, leading,
for $\vcr \in \overline{\tau_0}$, to
\begin{equation}
\psi(\vcr) = {\sf Y}(\vnr_0)^T {\sf R}^0(r_0) {\sf C}^0 \; ,
\end{equation}
for some unknown coefficients ${\sf C}^0$, with the difference being that
the ${\sf R}^0(r_0)$ are integrated inwards
from suitable boundary conditions at infinity.

The use of an outer sphere leads to a modified secular equation. We shall
not reproduce the algebra here since it is a relatively
straightforward adaptation of the method
of Ref.~\cite{Na+al.86a}, using the formulas and approach we have
developed above.

The utility of the outer sphere is that it enables one to impose boundary
conditions directly on the solution of the Schr\"{o}dinger equation,
important in the case of, for example, long-range Coulomb potentials.
It is also useful to reduce the size of the interstitial region, which
is important in the MT version of the MSW method,
and in the non-MT version of Ref.~\cite{Na+al.86a},
it helps to reduce the effect of the interstitial
$T$ matrix. For our present approach maximizing the size of the atomic
spheres (and minimizing the size of the outer sphere) would tend to
produce smaller potential gradients at the sphere surfaces
and hence a more gently varying distorting potential $V_I$ (if the atomic
parts of $V_I$ were chosen to be smooth continuations of the true
interstitial potential). 

As mentioned in the previous section, the use of an outer sphere
is a way of avoiding the possibility that $V_I$
has bound states of its own. This may
be done by choosing the extra-molecular part of $V_I$ to be
asymptotic to some constant potential sufficently less than zero
that no bound states arise in the resulting new $V_I$.
Since the old $V_I$ must be bounded from below,
one obvious choice would be the minimum of this potential,
although it may not be, in general, necessary to set the new asymptotic
value as low as this. For particular cases some
experimentation may be required. The smooth
continuation of the new $V_I$ into the extra-molecular region may be
then set identical to the chosen asymptotic value beyond some large
radius, so that the resulting ${\sf q}_i$ would be identical to the
diagonal modified spherical 
Hankel function forms for all larger values of the radius.

One consequence of this new effective zero for the $V_I$ is that
between this value and the true energy zero of $V$ we must use the
standing-wave Green function related to $V_I$, in a manner similar to
the original version \cite{Johnso73a} of the MSW method. This is,
however, expected to be a straightforward modification of the usual
positive-energy formulas (\ref{eq:gfinal}), with the ${\sf q}_i(r_i)$
now asymptotic to ${\sf n}(k^{'}r_i)$ and $k^{'}$ calculated relative
to the new asymptote of $V_I$.


\section{NUMERICAL IMPLEMENTATION}

The main computational task for the solution of the multicenter scattering
problem is the setting up of the secular matrix which is then inverted to
solve the secular equation.
The essential ingredients for this are the radial
solution matrices ${\sf p}_i$, ${\sf q}^{(+)}_i$,
and ${\sf R}^i$ for each atomic center.
For these we require the multipole components at each center
of $V_I$ from $r_i = 0$ to $\infty$, and of $V$ from $r_i = 0$ to $b_i$.
These, however, need only be calculated once. The radial solutions must be
calculated for each energy required. With these solutions we can evaluate
the constant Wronskian matrices ${\sf M}_{i +}$ at $r_i = b_i$, which are
then inverted; and the matrices ${\sf K}^{i j}$ for which the surface
integrals may also be evaluated at the same radius.
From these requirements we see that
${\sf p}_i$ and ${\sf R}^i$ are needed over the range $r_i = 0$ to $b_i$,
and ${\sf q}^{(+)}_i$ from $r_i = b_i$ to $\infty$. (Infinity here may be
taken to be a suitably large radius.)

The solution of coupled radial Schr\"{o}dinger equations for
various types of potential has received considerable attention
over many years and has a sizeable literature
(see for example Ref.~\cite{Th+al.81afull}),
and, for our immediate purposes, we
may consider currently available numerical methods adequate.
Although it is known that there are situations where problems may arise,
for the types of potential we consider they appear not to occur.
Our own experience with the matrix Numerov method, for example,
in the form given in Ref.~\cite{Na+al.86a}, has thrown up no
obvious difficulties; and, for calculations where an independent
check is available, gives quite acceptable results.


The key to an accurate implementation of our present approach is the
choice of distorting potential $V_I(\vcr)$. It is important that this
potential be well behaved, avoiding discontinuities
and wide variations. Since it is constrained
to coincide with the true interstitial potential in that region,
the only freedom we have is in the choice of values inside the atomic
volumes. Here we should replace the true singular atomic potential
with a smooth continuation of the
interstitial potential which is finite everywhere.
This may be done in a variety of ways. One possibility,
which has been used in a slightly different context by the author,
is to replace, for $r_i < b_i$,
each multipole component of the true potential
$V$ by a polynomial, in $r_i$, of minimal degree,
which matches the true multipole component at $r_i = b_i$,
and (perhaps) some of its derivatives.








\section{DISCUSSION}

Our new approach to non-MT MSW theory results in formulas similar in
overall form to those of some previous versions,
and indeed should be thought of as a direct development of them.
Although the derivation of the new secular equation in particular
owes much to the approach of Ref.~\cite{Na+al.86a},
the final form of the result is more reminiscent of the original
MT-based version of Slater and Johnson \cite{Johnso73a}.
This may be understood if one notes that our distorting potential
corresponds to the constant potential of their method. Thus where
one sees free-electron waves (expressed in terms of spherical Bessel,
Neumann, and Hankel functions) in their formulas, one sees the
distorted waves ${\sf p}_i$ and ${\sf q}_i$ in ours. 
(In fact, the MT version of the method should result from our present 
algebra when $V_I$ is replaced by the constant average interstitital
potential.) We treat directly the multiple scattering in terms
of waves which are already being scattered in the distorting potential.
By contrast, and very loosely speaking, the method of
Ref.~\cite{Na+al.86a} adds as an extra step (via the interstitial $T$
matrix) the effect of the interstitial region potential
on the waves scattering from the atomic centers.

To enable development of the non-MT MSW theory
in the direction we have taken
it requires, however, elaboration of the formal machinery that makes
up the bulk of this present work. It seems likely that some of the
results derived here are new and
should prove useful beyond our immediate subject.
Firstly, the explicit general Green-function
expansions of Ref.~\cite{Foulis04a} are necessary to
implement the distorted-wave Lippmann-Schwinger formalism,
leading, in particular, to an explicit form (\ref{eq:chip}) for the
distorted wave itself. Further direct development of the material
of Ref.~\cite{Foulis04a} was required, leading to some useful
formulas, notably the important relation (\ref{eq:gpartd1d1})
between regular and irregular matrix solutions of the coupled
radial Schr\"{o}dinger equations (\ref{eq:coupradschreq}).
We then saw how one can use these results to develop a complete
(if skeletal) solution to the single-scattering
problem for a general noncentral potential, whose most important
result is the formula (\ref{eq:scatampmtxa})
for the matrix ${\sf A}$ in the
expression for the scattering amplitude.
Indeed, having been presented with such a solution
one might reasonably ask why it could not be applied directly to
the scattering problem of the multicenter potential $V$.
As is well known, the singularities of $V$ have only slowly convergent
spherical-harmonic expansions relative to a single center. To have
an accurate treatment of scattering in such a potential requires an
approach which is capable of combining individual accurate
solutions in the regions of the singularities. This, of course,
is the essential utility of the
multiple-scattering approach to the solution
of Schr\"{o}dinger's equation for multicenter potentials.

As mentioned in Ref.~\cite{Foulis04a} certain mathematical issues
relating to the derivation and manipulation of our Green-function
expansions have been glossed over, and thus must be counted among
the assumptions of our approach. The use of vectors and matrices of
infinite dimension and questions of convergence should be considered
in more depth.
Such issues are frequently bypassed in the
literature, it being noted that for practical purposes such
vectors and matrices are truncated at finite dimension
in real calculations.
Of more physical interest is the asymptotic behavior of the
general irregular radial solutions ${\sf q}_i$. The derivation of
the Green-function expansions in Ref.~\cite{Foulis04a} assumes that
it is possible to choose these solutions asymptotic to a diagonal
form. Indeed we assume here that this diagonal form is the
free solution appropriate to outgoing waves
of Eq.~(\ref{eq:qplusfree}).
In the absence of finite charge moments the potential will typically
decay exponentially and this would almost certainly be the case.
It may easily be shown that the potential multipole expansion is
dominated by the monopole component
at infinity, even in the presence of
a finite dipole or higher moment. Thus the potential matrix
of Eq.~(\ref{eq:potmx})
will be diagonally dominant.
The nature of the limiting behavior necessary may in fact be
seen from the expression (\ref{eq:chipprelim1})
in the derivation of the explicit form for the scattered wave 
in a given potential, and in the algebra
[at Eq.~(\ref{eq:greenfndefncheck})] which verifies
the defining equation for the general Green function.

It is possible to bypass the issue of the asymptotic behavior of
the irregular solutions by using the outer-sphere option. Here one
may set the distorting potential to a constant value
beyond some finite radius. In this region, as we mentioned above, the
general irregular solutions may be chosen identical to the free ones,
although once inside the finite radius they will, in general, 
deviate from them.
This can be the case even if the true potential in the extra-molecular
region contains charge monopole, dipole or higher components.

The second important set of results which enables our version
of the non-MT MSW theory is that related to the re-expansion of 
general regular and irregular solutions of the coupled radial
Schr\"{o}dinger equations. This is central to our approach and we
have devoted a significant fraction of this present work to a
careful presentation of the derivations,
returning more than once to particular points,
and including alternative derivations of some formulas.
As mentioned in the section on bound states, and illustrated in
Eq.~(\ref{eq:chipprelim1}), the particular form of the
matrix ${\sf D}(k;\vcR_{ij})$ associated with translations
is dependent on the asymptotic form of ${\sf q}$. It is
interesting that this matrix, which has been encountered before in
MSW theory, has a more general function than previously appreciated.


\section{CONCLUSIONS}

We have presented here a new approach to multiple-scattering theory for
general potentials, based on the distorted-wave (two-potential)
Lippmann-Schwinger formalism. (We propose to refer to it therefore as the
``DW-MS'' method.) Our method makes use of the familiar partition of space 
into non-overlapping atomic spheres and a remaining interstitial region;
splits the multicenter molecular potential into a well behaved
background distorting potential and a singular atomic part; and uses some
newly developed Green-function expressions and re-expansion formulas to
derive the new secular equations.

The central improvement offered by this approach is to avoid any volume
integrals, which will have significant benefits for the scaling behavior
of calculations, the most complicated parts of which are now the surface
integrals associated with the matrices ${\sf K}^{i j}$. In fact, the
possibility arises, perhaps by taking the limit of the integrals as
$r_i \rightarrow 0$, that even these may be reduced to one-dimensional 
integrations. Calculation of the effects of the varying interstitial
potential appears implicitly via the irregular solutions
${\sf q}_i^{(+)}$ of the coupled radial Schr\"{o}dinger equations
for the distorting potential $V_I(\vcr)$. Another attractive feature 
of the approach is that the above calculational tasks require only
numerical methods that are already well tried and tested in related
approaches.

It may be noted that, in a formal sense at least, the separation between
geometry and scattering has been recovered. In reality of course this is
not quite true since there is a connection between the distorting
potential and the scattering centers. However, it may be possible to
exploit this in an approximate way to reduce the computational burden
in the context of geometry searches and total energy minimization.


\begin{acknowledgments}
The author would like to thank R.~F.~Pettifer, C.~R.~Natoli, and
N.~A.~M.~Pipon for some interesting discussions and encouragement.
\end{acknowledgments}

\bibliography{ppr022-dlf}

\newpage

\end{document}